\documentclass[aps,reprint,pra]{revtex4-1}
\pdfoutput=1
\usepackage{graphicx,amsmath,bm,multirow,xspace}

\usepackage{longtable,dcolumn}
\usepackage{subfigure,xcite}
\newcolumntype{d}[1]{D{.}{.}{#1}}

\usepackage{xcolor,ulem}

\newcommand{\rst}{\bgroup\markoverwith{\textcolor{red}{\rule[0.5ex]{2pt}{0.8pt}}}\ULon}

\newcommand{\homo}{\ensuremath{\varepsilon_{\text{HO}}}\xspace}
\newcommand{\ef}{\ensuremath{\varepsilon_F}\xspace}

\newcommand{\vs}{\ensuremath{v_s}}
\newcommand{\br}{\ensuremath{\bm{r}}}
\newcommand{\vws}{\ensuremath{V_{\text{prim}}}\xspace}

\begin{document}

\title{Classical turning surfaces in solids:\\
When do they occur, and what do they mean?}

\author{Aaron D. Kaplan}
  \email{kaplan@temple.edu}
\affiliation{Department of Physics, Temple University, Philadelphia, PA 19122}
\author{Stewart J. Clark}
\affiliation{Centre for Materials Physics, Durham University, Durham, DH1 3LE, United Kingdom}
\author{Kieron Burke}
\affiliation{Departments of Chemistry and Physics, University of California, Irvine, CA 92697}
\author{John P. Perdew}
\affiliation{Departments of Physics and Chemistry, Temple University, Philadelphia, PA 19122}

\date{\today}


\begin{abstract}

Classical turning surfaces of Kohn-Sham potentials, separating classically-allowed regions (CARs) from classically-forbidden regions (CFRs), provide a useful and rigorous approach to understanding many chemical properties of molecules. Here we calculate such surfaces for several paradigmatic solids. Our study of perfect crystals at equilibrium geometries suggests that CFRs are absent in metals, rare in covalent semiconductors, but common in ionic and molecular crystals. A CFR can appear at a monovacancy in a metal. In all materials, CFRs appear or grow as the internuclear distances are uniformly expanded. Calculations with several approximate density functionals and codes confirm these behaviors. A classical picture of conduction suggests that CARs should be connected in metals, and disconnected in wide-gap insulators. This classical picture is confirmed in the limits of extreme uniform compression of the internuclear distances, where all materials become metals without CFRs, and extreme expansion, where all materials become insulators with disconnected and widely-separated CARs around the atoms.

\end{abstract}

\maketitle

\section{Introduction}

The most basic property of an ordered solid is whether or not it is metallic \cite{MF36,K64,K68}. The Sommerfeld free electron model of metallic conduction \cite{SF31}, which involves quantum mechanics only via a Fermi distribution of velocities, assumes a homogeneous system (uniform electron gas), but we wish to understand the effect of inhomogeneity. A simple classical picture of conduction is to consider an electron of energy $\varepsilon$ in a single-particle effective potential, $v_{\text{eff}}(\br)$. If $\varepsilon > v_{\text{eff}}(\br)$ everywhere, this classical electron will move forever throughout the solid (or at least as far as its mean free path will allow), and the solid should be a metal. On the other hand, if the only classically allowed regions are disjoint regions bound to atoms, the solid should be strongly insulating. Unlike a classical electron, a quantum electron can tunnel into a classically-forbidden region.

The standard modern theory of conduction (for ordered solids) is that of Bloch bands, with insulators having filled bands below finite gaps in the spectrum \cite{am}. At first glance, this appears to have little in common with the simple classical picture given above. But quantum theories derive from classical theories, and are connected to quantum mechanics via semiclassical approximations using classical trajectories. Consider what happens to a standard band structure as $\hbar \to 0$ keeping the Fermi energy fixed. For energies above the maximum of the potential everywhere, the bands become more free-electron like, as the inhomogeneity in the potential becomes less relevant. On the other hand, for energies below the maximum, the band becomes narrower and more localized as $\hbar$ shrinks. The importance of turning points to semiclassical (and density functional) approximations was prefigured in the cartoon of Fig. 1 of Ref. \cite{ell08}.

The Kohn-Sham (KS) potential \cite{KS65} is the scalar potential that, acting on non-interacting electrons, yields a ground-state electron density equal to that of the real system. While not a physical observable, the KS potential is extremely useful as an interpretive tool. Inspired by earlier work that used the ``potential acting on an electron in a molecule'' (PAEM) \cite{yang1997,yang1998}, Ospadov {\it et al.} \cite{ospa2018} recently created a ``periodic table of nonrelativistic classical turning radii'' using the KS turning surface of the highest occupied KS orbital, defined as those points satisfying
\begin{equation}
  \vs(\br)= \homo, \label{eq:ts}
\end{equation}
where $\vs(\br)$ is the KS potential, and \homo is the energy of the highest occupied orbital (the Fermi energy \ef in a metal). They demonstrated that a classical turning surface could characterize bond types in molecules numerically and visually \cite{ospa2018}. At equilibrium geometries, covalent bonds as in N$_2$ have fused (roughly ellipsoidal) turning surfaces, ionic bonds as in NaCl often have seamed surfaces, hydrogen bonds as in (H$_2$O)$_2$ have necked surfaces, and van der Waals bonds as in Ne$_2$ have bifurcated surfaces (with each part nearly spherical). The ratio of an equilibrium bond length to the sum of its atomic radii is roughly 0.5 for a covalent bond, 1.0 for an ionic or hydrogen bond, and 1.5 for a van der Waals bond. More recently, Gould {\it et al.} \cite{gould2020} found that the classical turning surface of H$_2^+$, which is approximately ellipsoidal at the equilibrium bond length, bifurcates when the bond length is stretched to about twice the turning radius of one dissociation product H$^{+0.5}$ (rigorously the same as the turning radius of a neutral hydrogen atom, $\approx 1.06$ \AA{}). Neutral atoms other than hydrogen typically have one or more electrons in the classically-forbidden region (CFR) outside their turning surfaces \cite{schwprog}. Earlier, Ref \cite{burke16} had noted that a CFR emerges within the local density approximation (LDA) in stretched H$_2$ very near the Coulson-Fisher point, signaling the onset of strong correlation as the bond grows.

A turning surface in position space should not be confused with a Fermi surface in wavevector space. The turning surface defined here is the intersection in position-space of the KS potential with the Fermi level. If $\ef > \vs(\br)$ everywhere, there is no turning surface, whereas the Fermi surface is always well-defined. One could also define a turning surface in terms of the chemical potential $\mu \geq \ef$ \cite{pplb}, which differs from \homo for non-metals, but using \homo in Eq. \ref{eq:ts} is more practical and useful.

Here we present calculations of KS turning surfaces for a variety of simple solids. Our calculations are at the LDA and generalized gradient approximation (GGA) level of exchange-correlation approximations, which usually yield close approximations to more precise KS potentials in molecules (as both KS potential and \homo are typically too shallow by about the same amount). In Kohn-Sham density functional theory (KS DFT) \cite{KS65}, the KS potential $\vs(\br)$ is a multiplicative operator. In generalized KS theory, the exchange-correlation potential of a meta-GGA or a hybrid functional (using the Hartree-Fock exchange energy) is a non-multiplicative operator, but can be replaced \cite{ospa17} by the local one needed to define a classical turning surface. One can apply all the concepts of Ref. \cite{ospa2018} to analyze bonding in solids from a chemical viewpoint, but here we focus on the most elementary property of materials: are they metallic? In our classical conduction argument above, the effective potential is clearly the KS potential, and the most energetic electron is the highest-occupied level. If \homo is higher than the maximum value of the KS potential, there are no classical Fermi-energy turning surfaces and the system ought to be metallic. If not, and if \homo is so low that the classically-allowed regions are disconnected, the system ought to be insulating (with a wide gap). We expect semiconductors to lie somewhere in between these extremes.

This work discusses classical turning surface analogs and semiclassical interpretations of them for a variety of simple solids. Section \ref{sec:compmetd} describes the computational tools used to extract and analyze the KS potential for metals, as presented in Section \ref{sec:mets}, and band insulators presented in Section \ref{sec:insuls}. Special attention is paid to the roles of strain and defects in forming CFRs within solids. Section \ref{sec:per_trends} discusses how the CFR can be used to predict conduction properties. Section \ref{sec:connec} discusses the role of CFR connectedness in determining the conductive properties of solids. The Supplemental Materials section contains additional data.

To interpret our results correctly, we point out the following crucial points concerning gaps. It has long been known that the KS gap, i.e., the bandgap of the exact KS potential, does not match the true (fundamental charge) gap \cite{pplb}, and typically underestimates it. The KS gaps of strictly semilocal approximations like LDA or GGAs are typically close to the exact KS gap \cite{per85,per86,GMR06}, and thus are often substantially less than the fundamental gap. Hybrid functionals and meta-GGAs yield larger gaps when treated in a generalized KS scheme \cite{per17}. When lattice parameters are stretched well beyond equilibrium, semilocal functionals may produce broken-symmetry solutions of lower energy, as is well-known in the paradigmatic case of stretched H$_2$ \cite{PSB95}, but does not occur (at least for finite systems) with the exact functional. As all calculations in this paper use only semilocal functionals, they are in the KS scheme, yield gaps that are smaller than fundamental gaps, and can break symmetry.


\section{Computational Methods \label{sec:compmetd}}

All ensuing calculations were performed with either the Vienna {\it ab initio} Simulation Package (VASP) \cite{Kresse1993,*Kresse1994,*Kresse1996,*Kresse1996a}, or the Castep code \cite{castep1,castep2}, or both. All GGA calculations used the Perdew-Burke-Ernzerhof GGA \cite{pbe}, and all LSDA calculations used the Perdew-Zunger parameterization of the uniform electron gas correlation energy \cite{perd1981}. The calculations in VASP were performed with a cutoff energy of 800 eV, a $\Gamma$-centered mesh of spacing 0.08 \AA{}$^{-1}$, energy convergence of $10^{-6}$ eV, and stress convergence at $10^{-3}$ eV/\AA{}. To determine equilibrium geometries in VASP, for metals, first-order Methfessel-Paxton smearing with parameter of 0.2 was used, and for insulators, the Bl\"ochl tetrahedron method was used. VASP's internal methods were used to determine the relaxed cell volume. In Castep, a density-mixing algorithm was used to reach self-consistency, and geometries were determined with a BFGS (Broyden-–Fletcher-–Goldfarb-–Shanno) energy minimization scheme with the finite basis set corrected for stress \cite{fp-stress}. After relaxation, a calculation at the equilibrium volume using the Bl\"ochl tetrahedron method was performed to accurately determine the density of states. Accurate \cite{OTF} PAW on-the-fly pseudopotentials were used throughout. Tables \ref{tab:al_main_pbe_v_raw_dat} to \ref{tab:xe_group_18_pbe_c_raw_dat} (in the Supplemental Materials) present all raw data; machine readable data will be made available upon reasonable request.

For monolayers, a 45 $\times$ 45 $\times$ 1 $\bm k$-point grid was used in conjunction with the Bl\"ochl tetrahedron method. All other parameters remain the same from bulk calculations. The $c$ direction was padded with 30 \AA{} of vacuum region to reduce interactions between image monolayers.

In density functional plane-wave codes, the densities and potentials are stored on a uniform grid $\bm R$, the dimensions of which are determined by the size of the unit cell and the plane-wave cutoff energy. Acceptable convergence of the total energy relies on suitable convergence of the potentials and densities on this grid. The values of $\vs(\bm R)$ are obtained from this grid. In core regions, the true potential is much deeper than the pseudopotential, so these are classically allowed. Thus the PAW pseudopotential core regions were excluded from the CFR (frozen-core pseudopotentials were used in both VASP and Castep). The self-consistent electronic eigenstates give \homo (the Fermi energy \ef in a metal), and the regions where $\homo - \vs(\bm R) < 0$ define the CFR. We assign equal volume to each point relative to the primitive cell, as the real-space mesh is uniform. Suppose there are $N_{\text{prim}}$ total real-space mesh points in the  cell, and let the volume of the primitive cell be $\vws$. Then the volume of any point is $ \vws/N_{\text{prim}}$. If there are $N_{\text{CFR}}$ points at which $\homo - \vs(\br) < 0$, the volume of the CFR is
\begin{equation}
  V_{\text{CFR}} = \vws N_{\text{CFR}}/N_{\text{prim}}.
\end{equation}
The dimensionless, ``fractional volume'' of the CFR, which will be used in the ensuing figures and fits, is defined as
\begin{equation}
  v \equiv V_{\text{CFR}}/\vws = N_{\text{CFR}}/N_{\text{prim}},
\end{equation}
the number of real-space mesh points within the CFR relative to the total number of mesh points in the primitive cell.

As the fractional CFR volume $v \to 0$, our method requires ever finer real- and reciprocal-space meshes to resolve $v$. This need is limited by the resolution determined by the plane-wave cutoff energy. Our data for $v \ll 1$ will necessarily be more noisy than for larger values of $v$. Despite this, we show {\it a posteriori} that reasonable fits to $v(\vws)$ may be found.

Each code uses differently-generated pseudopotentials with different optimal basis set cutoff energies (and hence pseudopotential grid sizes, etc.), different energy minimization schemes, and different Brillouin zone integration methods. To ensure that our method is not dependent upon the numerical methods of a particular code, we have verified that the Castep and VASP results are consistent.

\section{Opening CFRs in metals \label{sec:mets}}

\begin{figure}
  \centering
  \includegraphics[width=\columnwidth]{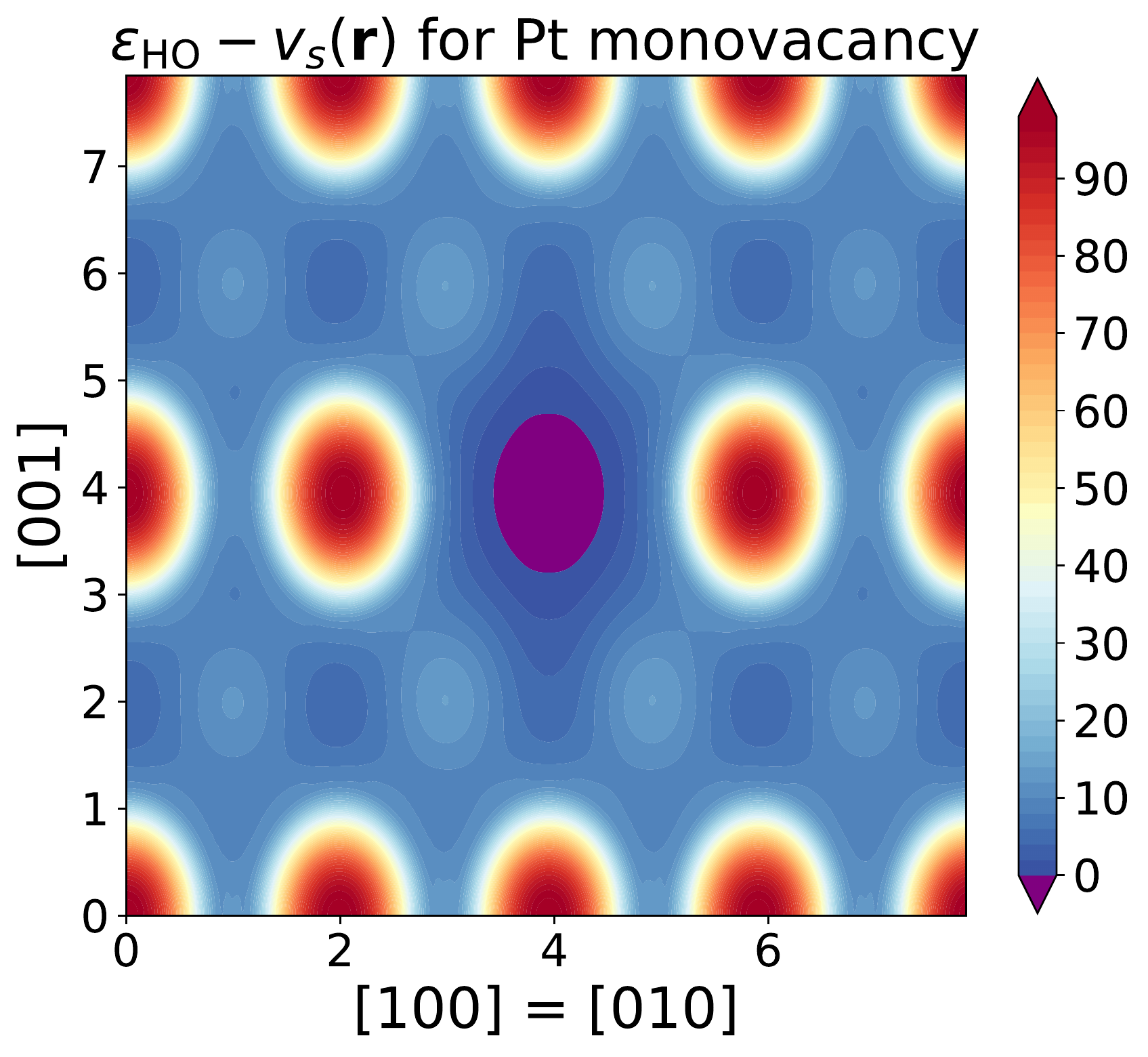}
  \caption{A contour plot of $\homo - \vs(\br)$ as calculated with PBE along the [110] (conventional cubic indices) direction in the Pt monovacancy supercell. The CFR (purple) surrounds the defect, supporting the conjecture that the formation of a defect is accompanied by the formation of an internal curved surface. Regions within the PAW pseudopotential core radii are only included here to make the image clearer. For an analogous figure in Si, refer to \ref{fig:sipot} in the Supplemental Materials. \label{fig:ptmv}}
\end{figure}

As we see in Table \ref{tab:pbemets}, no CFR is present in certain defect-free metals (Al, Cu, and Pt) at their equilibrium geometries. This is in line with our initial hunch, but does not extend to metals with monovacancies. A Pt supercell with a monovacancy defect harbors a small CFR; plotting this CFR in Fig. \ref{fig:ptmv}, we see that the classically-forbidden region encapsulates the center of the vacancy perfectly. Relaxation of the supercell volume was performed two ways: direct minimization of the stress tensor, and keeping the supercell volume fixed to the bulk volume while allowing ion positions to change.

The vacancy defect formation energy can be recast as the energy needed to create a curved surface within a solid \cite{perd91}. The localization of the CFR to the vacancy region is a clear manifestation of this. Carling {\it et al.} \cite{carl00} found that the LDA is more accurate than GGAs for the Al monovacancy formation energy, in line with earlier results \cite{cons08} for the jellium surface energy. They also found a very low electron density near the center of the vacancy, and large Friedel oscillations around it, consistent with a CFR near the center. Large voids and exterior surfaces would also give rise to extensive CFRs in any material.

The definition of the monovacancy volume given in Carling {\it et al.} differs from ours. Their method used the liquid drop model of jellium from Ref. \cite{perd91} to extract the vacancy's volume from the vacancy defect formation energy. This method will generally yield larger volumes than the corresponding CFR volumes.

\begin{figure}
  \centering
  \includegraphics[width=\columnwidth]{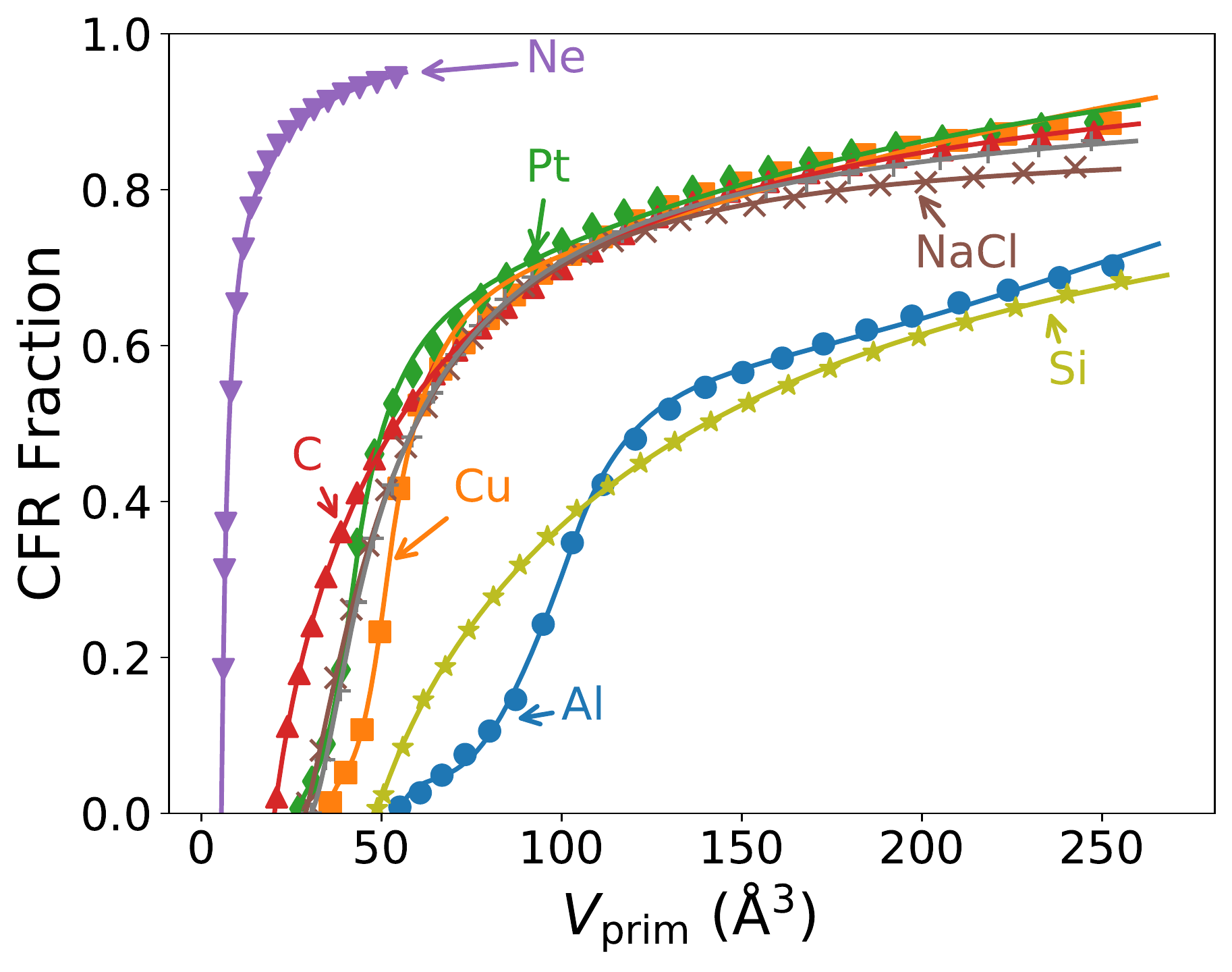}
  \caption{Emergence of PBE CFRs in Al (blue closed circles), Cu (yellow squares), Pt (green diamonds), C (red point-up triangles), Ne (purple point-down triangles), NaCl (brown crosses), and Si (olive stars) as a function of the primitive cell volume. All lines are fits given in Table \ref{tab:fits}. As Al, Cu, C and Si have no CFR at their relaxed lattice parameters, each lattice must be stretched to introduce a CFR. Conversely, Ne and NaCl must be compressed to eliminate their CFRs; for completeness, the full NaCl curve is presented here. The curve for spin-unpolarized NiO is almost identical to that of NaCl (see Fig. \ref{fig:pbe_dimensionless} of the Supplemental Materials). The LSDA curves are very similar, see Fig. \ref{fig:allstrn} in the Supplemental Materials. For PBE curves as calculated in Castep, see Fig. \ref{fig:pbe_comp} in the Supplemental Materials. \label{fig:pbestrn}}
\end{figure}

It is natural then to ask how far a metal needs to be stretched before a CFR emerges. In Fig. \ref{fig:pbestrn}, we plot the fractional volumes of the LSDA and PBE CFRs as a function of the primitive cell volume. To find accurate critical primitive cell volumes $V_c$ for the emergence of a CFR, we perform a least squares fit to
\begin{equation}
  v(\vws,\bm c) = \Theta(\vws-V_c) \sum_{m=0}^3 c_m \left(\frac{V_c}{\vws} \right)^m \label{eq:fitfn}
\end{equation}
where $\vws$ is the primitive cell volume, $0 \leq v < 1$ is the fractional CFR volume, $\Theta$ is a step function, and the dimensionless $\bm c \equiv (c_0,c_1,c_2,c_3)$ are derived from least-squares fit parameters. Note that $\sum_{m=0}^3 c_m = 0$, and $\partial v/\partial \vws > 0$ for $\vws - V_c = 0^+$. Fit parameters for PBE and LSDA data, as calculated in VASP, are listed in Table \ref{tab:fits}. For the fit parameters of PBE data as calculated in Castep, refer to Table \ref{tab:vac} in the Supplemental Materials.

Despite the possibility of noisy data at small CFR fractions, $0 < v \ll 1$, no lower cutoff on the VASP data was needed in the fitting process. The Castep data required a cutoff of $v_c = 0.01$ (for which any data with $v \leq v_c$ was taken to have $v = 0$ instead), and a few elemental solids (Mg, Sr, Ba, Ra) required a higher cutoff, $v_c = 0.05$.

\begin{table}
  \centering
  \begin{ruledtabular}
    \begin{tabular}{p{1.2cm}p{1cm}p{1cm}p{1cm}p{1cm}p{1cm}}
      Solid (structure) & $\homo - v_s^{\text{max}}$ (eV) & CFR fraction & \vws (\AA{}$^3$) & Lattice Const. (\AA{}) & Expt. Lattice Const. (\AA{}) \\  \hline
      \multirow{2}{1.2cm}{Al (fcc)} & 5.75 & 0\% & 16.48  & 4.04  & 4.02 \\
      & (5.94) & & (15.81) & (3.98) & \cite{sun2011} \\ \hline
      \multirow{2}{1.2cm}{Cu (fcc)} & 5.56 & 0\% & 12.01  & 3.63 & 3.59 \\
      & (6.04) & & (10.94) & (3.52) & \cite{sun2011} \\ \hline
      \multirow{2}{1.2cm}{Pt (fcc, bulk)} & 4.76 & 0\% & 15.61 & 3.97  & 3.91 \\
      & (5.04) & & (14.90) & (3.90) &  \cite{haas2009} \\ \hline
      \multirow{4}{1.2cm}{Pt monovacancy (fcc)} & -1.18  & 10.9\%  & 15.39  & 3.95  & 3.91 \\
      & (-1.00) & (5.2\%) & (14.68) & (3.89) &  \\
      & -1.29 & 12.1\% & 15.61  & 3.97 & 3.91 \\
      & (-1.29) & (6.4\%) & (14.90) & (3.90) &
    \end{tabular}
    \caption{PBE and LSDA (parenthesized when different) values for the classically-forbidden regions (as a percentage of the total cell volume) and the relaxed primitive cell volumes and lattice constants in select metals. The percent volume is taken with respect to the primitive unit cell (percent volume per atom). For the first set of Pt monovacancy results, the cell volume and ion positions were relaxed; for the second set, the volume was fixed to the bulk value, and the ion positions were relaxed. Both sets of calculations used 31 ions in the supercell. \label{tab:pbemets}}
  \end{ruledtabular}
\end{table}

The form of Eq. \ref{eq:fitfn} is selected because it makes $v$ tend to zero as \vws tends to $V_c$ from above, and to a finite value ($c_0$) as \vws tends to infinity. (A perfect fit over the whole range $\vws > V_c$, not needed here, would require $c_0 = 1$.) As the PAW core region is classically-allowed in an all-electron approach, there will always exist a classically-allowed region in this type of calculation. As our method has lower resolution for $v \ll 1$, we expect our fitted values of $V_c$ to be estimates of the ``true'' values. $V_c$ was determined by a root-finding algorithm. When possible, the value of $V_c$ was constrained to lie between the largest tabulated value of \vws for which $v = 0$, and the smallest tabulated value of \vws for which $v> 0$. When that was not possible, a tolerance of 3\% was afforded, which we note in Table \ref{tab:fits} as well.

\begin{table}
  \centering
  \begin{ruledtabular}
    \begin{tabular}{p{1cm}ccccccc}
      Solid (struc.) & DFA & $c_0$ & $c_1$ & $c_2$ & $c_3$ & $R^2$ & $V_c$ (\AA{}$^3$) \\ \hline
       & LSDA & 0.59 & -1.32 & 0.90 & -0.17 & 0.0008 & 48.08 \\
      Al & & 0.67 & 1.25 & -7.22 & 5.96 & & 71.18 \\
      (fcc) & PBE & 5.03 & -17.31 & 20.16 & -7.88 & 0.0011 & 54.68 \\
      & & 1.71 & -8.48 & 22.46 & -21.36 & & 103.65 \\ \hline
       & LSDA & 7.37 & -24.41 & 27.18 & -10.14 & 0.0002 & 29.69 \\
      Cu & & 1.04 & -1.47 & 2.31 & -3.14 & & 47.31 \\
      (fcc) & PBE & 12.97 & -42.29 & 46.44 & -17.12 & 0.0051 & 35.02 \\
      & & 1.25 & -3.40 & 7.95 & -7.35 & & 51.29 \\ \hline
       & LSDA & 7.37 & -23.91 & 26.14 & -9.60 & 0.0002 & 25.61 \\
      Pt & & 1.01 & -1.41 & 1.82 & -2.30 & & 39.01 \\
      (fcc) & PBE & 7.26 & -23.32 & 25.35 & -9.29 & 0.0027 & 26.78 \\
      & & 1.13 & -2.71 & 6.17 & -6.26 & & 42.36 \\ \hline
      C & LSDA & 1.01 & -1.55 & 0.81 & -0.27 & 0.0006 & 23.16 \\
      (ds) & PBE & 1.02 & -1.79 & 1.27 & -0.50 & 0.0004 & 20.26 \\ \hline
      Ne & LSDA & 0.97 & -0.27 & -0.25 & -0.44 & 0.0068 & 5.85 \\
      (fcc) & PBE & 1.02 & -0.84 & 1.04 & -1.22 & 0.0015 & 5.54 \\ \hline
      NaCl & LSDA & 0.89 & -0.23 & -1.86 & 1.20 & 0.0002 & 30.70 \\
      (rs) & PBE & 0.86 & -0.09 & -1.90 & 1.13 & 0.0002 & 28.88 \\ \hline
      Si & LSDA & 0.92 & -1.34 & 0.52 & -0.10 & 0.0005 & 53.82 \\
      (ds) & PBE & 0.96 & -1.66 & 1.14 & -0.44 & 0.0001 & 48.59 \\ \hline
      NiO & PBE & -0.14 & 4.45 & -8.49 & 4.17 & 0.0004 & 29.56 \\
      (rs) & & 0.94 & -0.68 & -0.40 & -0.07 & & 48.29 \\
    \end{tabular}
    \caption{Parameters for the fit functions presented in Figs. \ref{fig:pbestrn} and \ref{fig:allstrn} (in the Supplemental Materials). The density functional approximation (DFA) column refers to either LSDA or PBE in VASP; for Castep fits, see Table \ref{tab:vac} of the Supplemental Materials. The $\bm c$ parameters are dimensionless. $R^2$ is the sum of square residuals. $V_c$ is the predicted critical primitive cell volume for onset of a CFR. For metals, the first (second) line gives the parameters $\bm c$ for $\vws < V_0$ ($\bm d$, for $\vws > V_0$, and $V_0$ is given in lieu of $V_c$). The fitted LSDA $V_c$ for Ne is too large (i.e., it is larger than the smallest tabulated value of \vws for which $v(\vws)>0$); all other fitted values of $V_c$ are within the correct bounds. Here, ``ds'' refers to diamond structure and ``rs'' to rock salt structure. NiO is treated as spin-unpolarized. \label{tab:fits}}
\end{ruledtabular}
\end{table}

Many of the metals presented here exhibit more complex \vws-dependence than the insulators, so we perform a piecewise fit
\begin{eqnarray}
  v_{\text{metal}}(\vws) &=& v(\vws,\bm c)\Theta(V_0 - \vws) \nonumber \\
  & & + v(\vws,\bm d) \Theta(\vws - V_0),
\end{eqnarray}
where both functions on the RHS are of the form of Eq. \ref{eq:fitfn}. To perform the fit, we chose a value of $V_0$ to model the point of inflection of the curve, and the fitting procedure detailed above was followed for $\vws \leq V_0$. We then required that $v(\vws)$ and $\partial v(\vws)/\partial \vws$ be continuous at $\vws = V_0$, fixing $d_0$ and $d_1$. A least squares fit was then performed to yield $d_2$ and $d_3$. $V_0$ was modulated to minimize the sum of square residuals, $R^2 = \sum_{\vws}|v^{\text{approx}}(\vws) - v(\vws)|^2$. To prevent over-fitting, the lowest value of $V_0$ for which $R^2 < 10^{-3}$ was deemed the optimal fit.

Let $a$ be the equilibrium lattice parameter for a given solid as given by Table \ref{tab:pbemets}. From the critical lattice parameters in Table \ref{tab:fits}, we see that a CFR opens in Cu at a lattice parameter of $1.43a$ for PBE (about 2.9 times the equilibrium volume). The CFR appears before the KS gap opens. The fits predict that a CFR in Al opens at $1.49a$ for PBE (about 3.3 times the equilibrium volume), also without a KS band gap opening. For Pt, the CFR opens at $1.20a$ (about 1.7 times the equilibrium volume), without a gap opening. By bandgap, we always mean the gap determined from our approximate band structure or density of states, rather than the fundamental gap.

Note also that the LSDA and PBE curves in Figs. \ref{fig:pbestrn} and \ref{fig:allstrn} for Al, Cu, and NaCl cross, whereas those for elemental insulators do not. For the elemental insulators, the difference between the LSDA and PBE curves is always of the same sign.


\section{CFRs in insulators \label{sec:insuls}}

The situation for insulators, as shown in Table \ref{tab:pbeinsul}, is more nuanced, and there are clear variations in the volumes of approximate CFRs found from different approximate exchange-correlation functionals. Our intuition that the presence of a CFR is accompanied by the opening of a band gap is not borne out.

\begin{table*}
  \centering
  \begin{ruledtabular}
    \begin{tabular}{p{1.5cm}p{1.5cm}p{3cm}p{3cm}p{2cm}p{2.5cm}p{2.5cm}}
      \multicolumn{2}{c}{Solid (structure)} & $\homo - v_s^{\text{max}}$ (eV) & CFR fraction & \vws (\AA{}$^3$) & Lattice Const(s). (\AA{}) & Expt. Lattice Const(s). (\AA{}) \\  \hline
      Si (ds) & & 0.91 (1.44) & 0\% & 40.89 (39.43) & 5.47 (5.40) & 5.42 \cite{sun2011} \\ \hline
      C (ds) & & 5.59 (6.62) & 0\% & 11.40 (11.04) & 3.57 (3.53) & 3.55 \cite{sun2011} \\ \hline
      \multirow{2}{3cm}{C (hex)} & relaxed & -3.06 (1.04) & 18.5\% (0\%) & 42.80 (34.45) & 2.47 (2.45) ($a$), 8.12 (6.65) ($c$) & 2.46 ($a$), 6.71 ($c$) \cite{zhao1989,truc1975} \\
       & expt. & -0.28 (0.92) & 1.0\% (0\%) & 35.25 & 2.46 ($a$), 6.71 ($c$) & 2.46 ($a$), 6.71 ($c$) \\ \hline
       \multirow{2}{3cm}{Ne (fcc)} & relaxed & -14.44 (-9.23) & 87.1\% (78.1\%) & 23.67 (14.39) & 4.56 (3.86) & 4.46 \cite{schw2009} \\
       & expt. & -14.20 (-10.90) & 86.3\% (86.3\%) & 22.24 & 4.46 & 4.46 \\ \hline
       \multirow{2}{3cm}{NaCl (rs)} & relaxed & -3.38 (-2.05) & 34.6\% (17.7\%) & 46.23 (40.90) & 5.70 (5.47) & 5.57 \cite{sun2011} \\
       & expt. & -3.06 (-2.30) & 29.6\% (22.4\%) & 43.18 & 5.57 & 5.57 \\ \hline
       MoS$_2$ (P6/mmc or $2H_b$) & & -4.24 (0.01) & 22.3\% (0\%) & 128.46 (101.94) & 3.18 (3.12) ($a$), 14.62 (12.07) ($c$), 3.12 (3.11) ($z$) & 3.16 ($a$), 12.29 ($c$), 3.17 ($z$) \cite{boker01} \\ \hline
       NiO (rs) & & 4.97 & 0\% & 18.01 & 4.16 & 4.17 \cite{tra06} \\
    \end{tabular}
    \caption{PBE and LSDA (parenthesized when different) values for the classically-forbidden regions, primitive cell volumes, and lattice constants of select insulators. For graphite, Ne, and NaCl, two sets of results are shown: the first at a relaxed PBE geometry, and the second at the experimental equilibrium geometry. The percent volume is taken with respect to the primitive cell (percent volume per atom). Here, ``ds'' refers to diamond structure, ``hex'' to simple hexagonal structure (with a four-point basis for graphite), and ``rs'' to rock salt structure. The layered structure of MoS$_2$ is itself a prototype for dichalcogenide structure, and is often referred to as the ``MoS$_2$ structure,'' or by its polytype $2H_b$ \cite{boker01}, or by its point group P6/mmc \cite{kan14}. The $a$ and $c$ parameters have the same meaning as in a simple hexagonal lattice, the $z$ parameter (sometimes called $2z$) is the spacing between neighboring sulfur layers. No LSDA calculation was performed for spin-unpolarized NiO. \label{tab:pbeinsul}}
  \end{ruledtabular}
\end{table*}

However in weakly interacting and van der Waals solids, like graphite and Ne, there are noticeable PBE CFRs. The small (1\%) PBE CFR volume in graphite (hexagonal C) at its experimental lattice constants reflects the semimetallic nature of this material. The PBE CFR in graphite lies between monolayers, just as one might expect for few-layer graphene. The CFR volume is nearly 20\% of the PBE equilibrium cell volume in graphite because PBE underestimates intermediate-range van der Waals interactions, and thus overestimates the equilibrium spacing. This fraction is reduced to 1\% when the experimental cell volume is used instead. The LSDA finds no CFR in graphite, which may be related to the LSDA's underestimation of equilibrium lattice constants. For the prototypical semiconducting layered material MoS$_2$, we see the same pattern. The LSDA underestimates the equilibrium $c$ lattice parameter, yielding no CFR. PBE dramatically overestimates the $c$ parameter, yielding a CFR encompassing 22.3\% of the primitive cell volume. Note that the LSDA and PBE are similarly accurate for the sandwich-layer thickness $z$ (the distance between neighboring layers of sulfur atoms).

Consider instead a monolayer of graphite or MoS$_2$. For these sheets, we use the bulk $a$ and $z$ lattice parameters found by relaxing the equilibrium cell volume. We find no CFR within the monolayer region for graphene or monolayer MoS$_2$ using both the LSDA and PBE. Thus, no in-plane CFR is present in graphene, and no in-sandwich CFR is found in monolayer MoS$_2$.

Crystalline NaCl, just like its molecular form \cite{ospa2018}, also has large PBE and LSDA CFRs. Because NaCl is a prototypical ionic solid, we expect that many other ionic crystals and more weakly-bound crystals at equilibrium will exhibit CFRs.

Referring back to Table. \ref{tab:fits}, we see that a CFR emerges in ds C when the lattice is stretched to $1.21a$ for PBE (about 1.8 times the equilibrium volume); for ds Si, a CFR is predicted to emerge at $1.06a$ for PBE (about 1.2 times the equilibrium volume). Thus it appears that PBE predicts the emergence of a CFR in an insulator when the lattice is stretched not much further past its equilibrium point. For both Si and C, the band gap is substantial even when the CFR begins to emerge.

The classical radius of the free Ne atom is 0.87 \AA{}, in both PBE \cite{ospaprog}, and with a more accurate Kohn-Sham potential \cite{ospa2018}, with a volume of 2.76 \AA{}$^3$. The experimental lattice constant is 4.464 \AA{} \cite{schw2009}, corresponding to a cell volume per atom of 22.24 \AA{}$^3$. The CFR predicted by Ref. \cite{ospa2018} is then $(22.24 - 2.76)/22.24 \approx 88$\% of the total cell volume, agreeing with the values in Table \ref{tab:pbeinsul}. A Ne atom in solid Ne at the equilibrium lattice constant is very similar to a free Ne atom.

In the same vein as for C and Si, we can compress the Ne lattice until the CFR vanishes, as seen in Fig. \ref{fig:pbestrn}. The Ne CFR is predicted to vanish at $0.62a$ for PBE. One might expect the bandgap to shrink as the CFR collapses, but the {\it opposite} is true. For the smallest lattice constant calculated here (2.85 \AA{}), the band gap is roughly 18.57 eV, compared to a gap of about 11.51 eV (11.45 eV) at the PBE equilibrium (experimental) lattice constant, consistent with previous work that used PBE to study phases of Ne under pressure \cite{he10}. Intuition suggests that the Ne CFR should not be fully suppressed before the classical turning surfaces between adjacent atoms just touch, at a nearest-neighbor separation of $2(0.87) = 1.74$ \AA{}, using the result from Ref. \cite{ospa2018}. This is substantially smaller than the nearest-neighbor spacing in crystalline Ne for which the PBE CFR is wiped out, $2.81/\sqrt 2 \approx 2.00$ \AA{}. Thus, unexpectedly, the critical lattice constant in Ne makes the nearest-neighbor distance noticeably greater than twice the turning radius of the free atom.

\begin{figure}[t]
  \centering
  \includegraphics[width=\columnwidth]{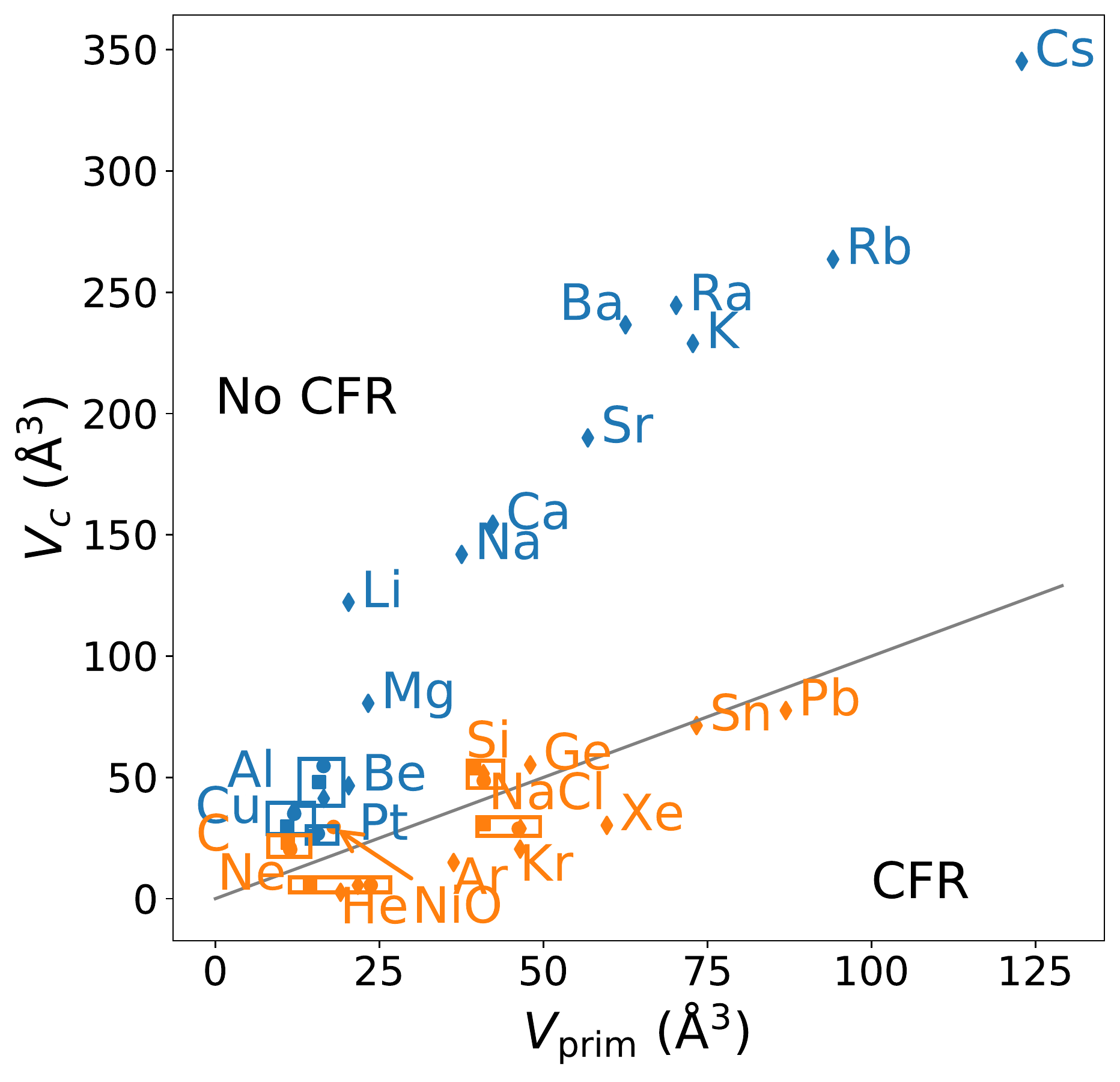}
  \caption{ Contrasting the amount of strain needed to induce a CFR for various solids. The grey line is $\vws = V_c$, so that solids lying above the grey line have no CFR at equilibrium, and solids below the line have a CFR at equilibrium. For the VASP data (circles for PBE and squares for LSDA), \vws is the equilibrium cell volume. For the Castep data (diamonds all with PBE), \vws is the equilibrium cell volume except for the rare gases, for which \vws takes the experimental value. Metals are shown in blue, and insulators in orange. Note that Sn and Pb were calculated in the cubic diamond structure ($\alpha$- or grey Sn), which are likely non-metallic phases. In particular, grey Sn has a 0.1 eV gap \cite{am}. NiO is treated as spin-unpolarized. \label{fig:vwsvc}}
\end{figure}

As the lattice is compressed, two competing effects determine the bandgap: the bands widen, reducing the gap; and the center of the conduction band is shifted upwards with respect to the center of the valence band, widening the gap. (For an example, see the silicon density of states at equilibrium and at a mild expansion, in Fig. \ref{fig:sidos} of the Supplementary Material.) This leads to a nontrivial (non-monotonic) dependence of the gap upon the lattice parameter.

\section{Periodic Trends \label{sec:per_trends}}

To get an overall sense of how well (or poorly) the classical turning surface yields information on conduction, Fig. \ref{fig:vwsvc} demonstrates one way to classify this, by plotting the fitted values of $V_c$ against the equilibrium \vws for various solids. The elemental solids beyond those emphasized in the main text are in Groups 1 (alkali metals), 2 (alkaline earth metals), 14 (Group IV, carbon group) and 18 (rare, inert, or noble gases). The parameters of the fit functions (Tables \ref{tab:group1}-\ref{tab:group18}), as well as full strain curves for these solids (Figs. \ref{fig:group1}-\ref{fig:group18}) can be found in the Supplementary Materials. The figure shows that the existence of a CFR at equilibrium comes close to classifying a solid as an insulator or metal. No metal has an equilibrium CFR, but some insulators need a small expansion to produce one.

\begin{figure}[t]
  \centering
  \includegraphics[width=\columnwidth]{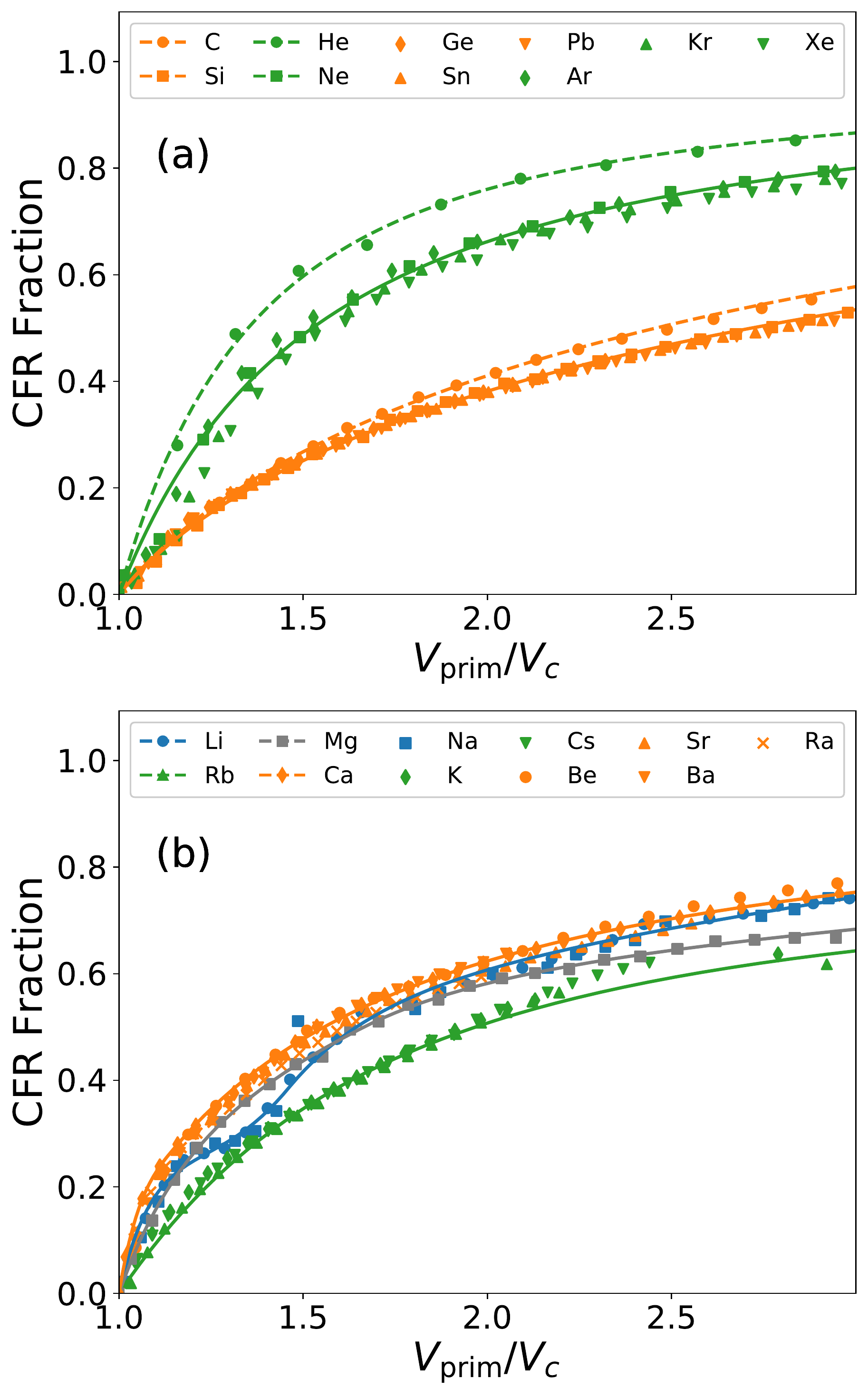}
  \caption{ Trends among groups of elements emerge when plotting the CFR fraction against the dimensionless $\vws/V_c$, where $V_c$ is fitted. All data shown here was calculated with PBE in Castep. Panel (a) is for the metals, and panel (b) for the insulators. \label{fig:per_trends}}
\end{figure}

Moreover, we can see very clear trends in the strain curves of elemental solids as one goes down a column of the periodic table. In Fig. \ref{fig:per_trends}(a), we plot the strain curves as a function of $\vws/V_c$, for elemental insulators. The noble gases all fall on one line, except for the lightest, He, while the carbon group elements fall on another, except for the lightest, C. Clearly, each group has its own characteristic curve, which differs from one group to another.

The alkali and alkali-earth metals show similar, but more complex behavior, as shown in Fig. \ref{fig:per_trends}(b). The green line is for the heavier alkalis, and the orange line is for the alkali-earths. Now the lighter two alkalis, Li and Na, are shown in blue, and clearly share a shape that is distinct from the later alkalis. They follow the alkali earth curve closely, except for a dip around $1.4V_c$. Moreover, Mg (in gray) is the odd one out of the alkali earths, rather than Be. For small strains, Mg behaves like all other alkali earths, but when greatly expanded, behaves more like an alkali.

Naturally, within a column of the periodic table, the critical CFR volume $V_c$ increases with atomic number, as shown in the Supplementary Materials. If we define the volume of a free atom as $V_{\text{at}} = 4\pi r_{\text{TS}}^2/3$, where $r_{\text{TS}}$ is the radius of the atom's classical turning surface from Ref. \cite{ospa2018}, then the ratio $V_c/V_{\text{at}}$ is of order 1 and seems to approach a column-dependent large-$Z$ limit (with $Z$ the nuclear charge, see Tables \ref{tab:group1}-\ref{tab:group18}). The first ionization energies of the atoms exhibit similar behavior \cite{cons2010}.

\section{CFR Connectedness \label{sec:connec}}

In the introduction, we described a (semi-) classical model of solids that defined metals and insulators by their turning surface properties. In this model, a metal would have a connected classically-allowed region (CAR), and an insulator would have a disconnected CAR.

\begin{figure}[ht]
  \centering
  \includegraphics[width=\columnwidth]{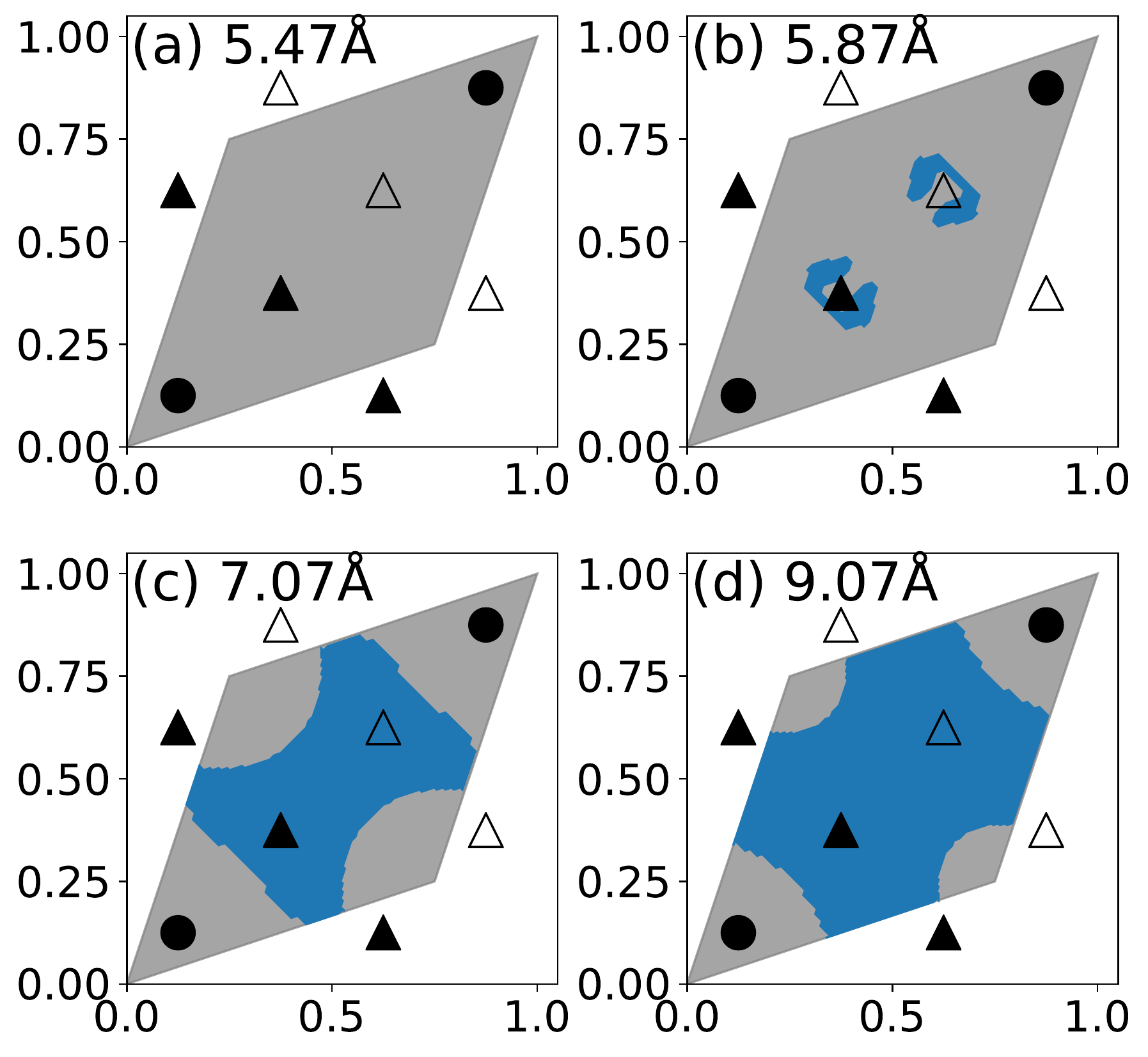}
  \caption{Evolution of the PBE CAR (grey) and CFR (blue) in ds Si as a function of the lattice parameter $a$ plotted along the plane $z=(x+y)/2$ (i.e., the plane containing the line joining the origin and the point $(a,a,a)$) within the primitive cell. Ions located in plane are labelled with a black circle, those above with a black triangle, and those below with an open triangle. The lattice constant $a$ is 5.47 \AA{} in (a), 5.87 \AA{} in (b), 7.07 \AA{} in (c), and 9.07 \AA{} in (d). \label{fig:si_cafr}}
\end{figure}

\begin{figure}[ht]
  \includegraphics[width=\columnwidth]{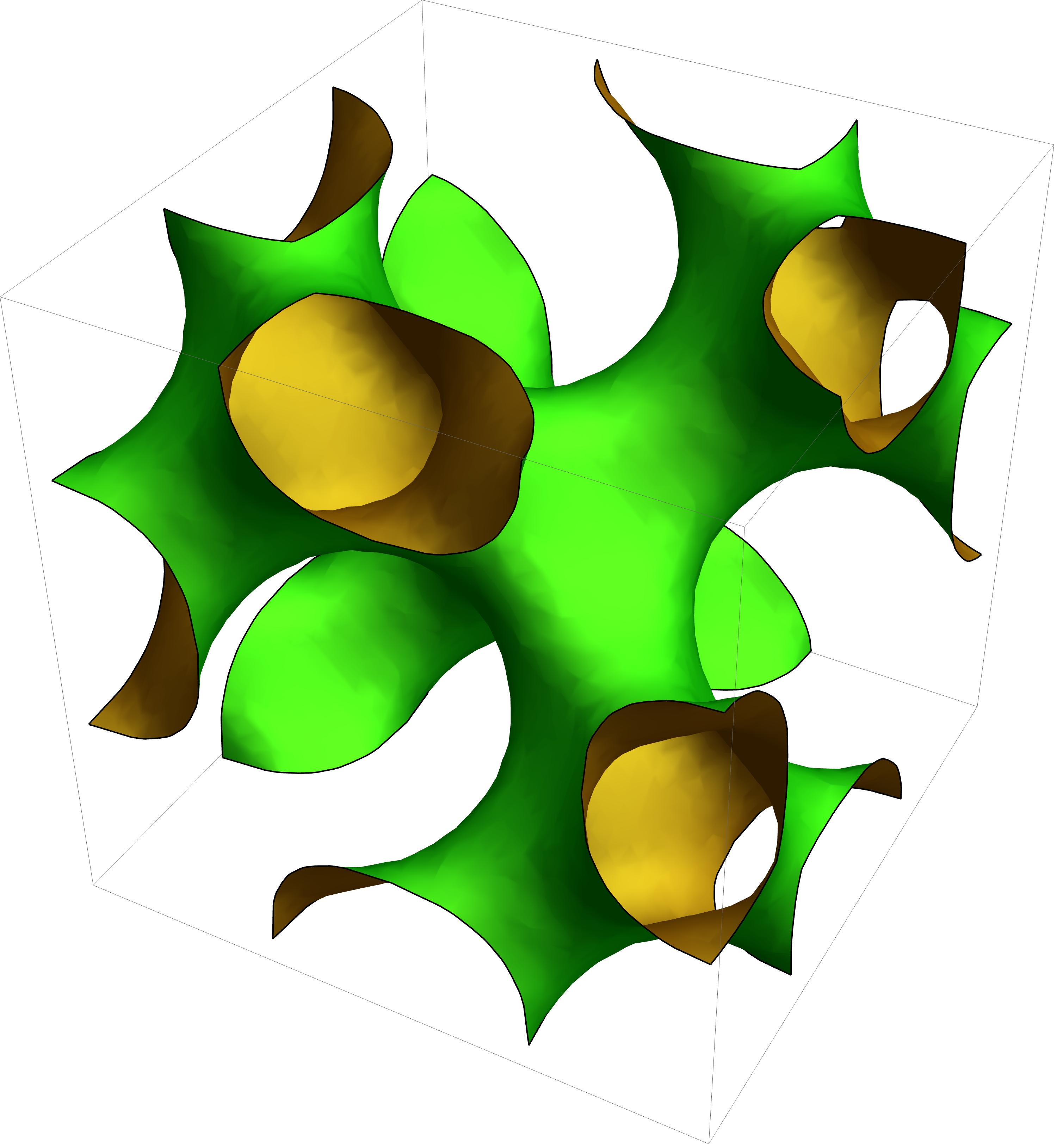}
  \caption{ The surface shows the CAR (outside, green) and CFR (inside, yellow) for silicon at 1.30$a$. Both regions are simultaneously fully connected. \label{fig:si_surf_plot}}
\end{figure}

In Figure \ref{fig:si_cafr}, we plot the evolution of the CFR and CAR in Si using PBE as a function of the lattice parameter. At equilibrium (panel (a)), there is no CFR. Just above $V_c$ (panel (b)), the CAR is clearly connected and the CFR is disconnected. As the lattice is stretched further (panel (c)), the CFR grows and connects. Under an even more extreme strain (panel (d)), the CFR dominates the primitive cell, but the CAR remains connected, albeit not simply. The bandgap increases from 0.55 eV at $a = 5.47$ \AA{} to 0.81 eV at $a = 5.80$ \AA{}, then decreases to 0.71 eV at $a = 5.87$ \AA{} before rapidly falling toward zero. In Fig. \ref{fig:si_surf_plot}, we show a three-dimensional view of the Si turning surface at $a = 7.11$ \AA{}. Both the CFR and CAR are simultaneously fully connected. The geometry of Fig. \ref{fig:si_surf_plot} is very nearly identical to that of Fig. \ref{fig:si_cafr}(c). To generate the plane of Fig. \ref{fig:si_cafr}(c) from Fig. \ref{fig:si_surf_plot}, one would make a diagonal cut from the front bottom left corner  to the rear upper right corner of the cell in Fig. \ref{fig:si_surf_plot}.

In the limit of extreme expansion, the CAR's are always disconnected and well-separated. Electron tunneling is inhibited through energy barriers that are wide or high. The bands will narrow to atomic levels. For a solid built from closed-shell atoms like Ne, the bandgap will tend to a non-zero and typically large value, while for a solid built up from open-subshell atoms like silicon, the bandgap will tend to a zero or small value (depending on whether the Kohn-Sham potential of the free atom is constrained to the symmetry of its external potential).

As the lattice is put under extreme expansive strain, we expect all solids to eventually transition to an insulating state, with disconnected and well-separated CAR's. In this limit, the bandgap can become large, small, or zero. If the lattice is compressed well below the equilibrium geometry, we expect any solid to eventually transition to a metallic state, with zero bandgap and a connected CAR.


\section{Conclusions}

For the solids studied here, our calculations found no CFRs for metals, large CFRs for wide-gap insulators, and the emergence of CFRs when small-gap semiconductors are mildly expanded. Since standard density-gradient expansions are derived for slowly-varying densities without CFRs, the absence of CFRs in metals at equilibrium suggests that generalized gradient approximations (like or beyond PBE) should work especially well for them.

A monovacancy in a metal can induce a CFR, and an expansive strain in any material can induce a CFR or increase its volume. Moreover, the emergence of a CFR does not necessarily accompany the opening or closing of a band gap.

The volume of a CFR is a function of the lattice strain. Both metals and band insulators without a CFR at their equilibrium geometry can be stretched to introduce a CFR. Those wider-gap insulators with a CFR at equilibrium can be compressed until the CFR vanishes. Layered materials may have a CFR at equilibrium if a density functional approximation tends to stretch the $c$ lattice parameter, as PBE does.

CFRs are also characteristic of perfect ionic and molecular crystals at equilibrium. Our analysis supports the conclusion that rare gas atoms in the crystalline phase are nearly free. Ionic crystals have large CFRs. We showed that graphite and MoS$_2$, where intermediate-range van der Waals interactions dominate between monolayers, have CFRs located solely between monolayers, and that their corresponding monolayers have no in-plane CFR. Our work demonstrates that weakly-bound solids tend to have prominent CFRs. Hydrogen-bonded crystals like ice, while not tested here, can be expected to have substantial CFR volume fractions, as suggested by Fig. 8 for the water dimer in Ref. \cite{ospa2018}.

The connectedness of a CFR seems to play a role in a system's conductivity. It was shown that the CFR in Si near the critical volume $V_c$ is disconnected. As the lattice is stretched further, the CFR grows, eventually subsuming much of the primitive cell. The Si bandgap closes very nearly at the same \vws that the CFR connects, indicating a semiclassical insulator-metal transition. A semiclassical picture suggests that a connected CAR and zero bandgap indicate a metallic state, found under extreme compression for any solid. As a corollary, disconnected and well-separated CAR's indicate an insulating state, and are found under strong expansion of the lattice.

Interacting quantum mechanical electrons can insulate through the Mott mechanism. We looked for CFRs in zero-gap spin-unpolarized NiO, a paradigm Mott insulator, but did not find one at the equilibrium lattice constant $a$. A CFR appeared at a lattice constant 1.18$a$ (see Tables \ref{tab:fits} and \ref{tab:pbeinsul}). Our PBE calculations for NiO at equilibrium confirmed that a gap appears when the spin symmetry is allowed to break to antiferromagnetic order.

This is the first work to attempt to classify CFRs in solids, and without doubt more inquiry is needed to determine if CFRs are hallmarks of other phenomena in solids.

\begin{acknowledgments}

ADK acknowledges the support of the Department of Energy (DOE), Basic Energy Sciences, under grant No. DE-SC0012575. SJC acknowledges Engineering and Physical Sciences Research Council support on grant EP/P022782/1. KB was supported by DOE under grant no. DE-FG02-08ER46496. JPP acknowledges the support of the National Science Foundation under grant number DMR-1939528.

\end{acknowledgments}


\bibliographystyle{apsrev4-2}
\bibliography{cfrbib}

\clearpage

\renewcommand{\thepage}{S\arabic{page}}
\renewcommand{\thesection}{S\arabic{section}}
\renewcommand{\theequation}{S\arabic{equation}}
\renewcommand{\thetable}{S\arabic{table}}
\renewcommand{\thefigure}{S\arabic{figure}}

\setcounter{page}{1}
\setcounter{section}{0}
\setcounter{equation}{0}
\setcounter{table}{0}
\setcounter{figure}{0}

\onecolumngrid

\section*{Supplemental Materials for \\
``Classical turning surfaces in solids:\\
When do they occur, and what do they mean?''}

\begin{figure}
  \includegraphics[width=\columnwidth]{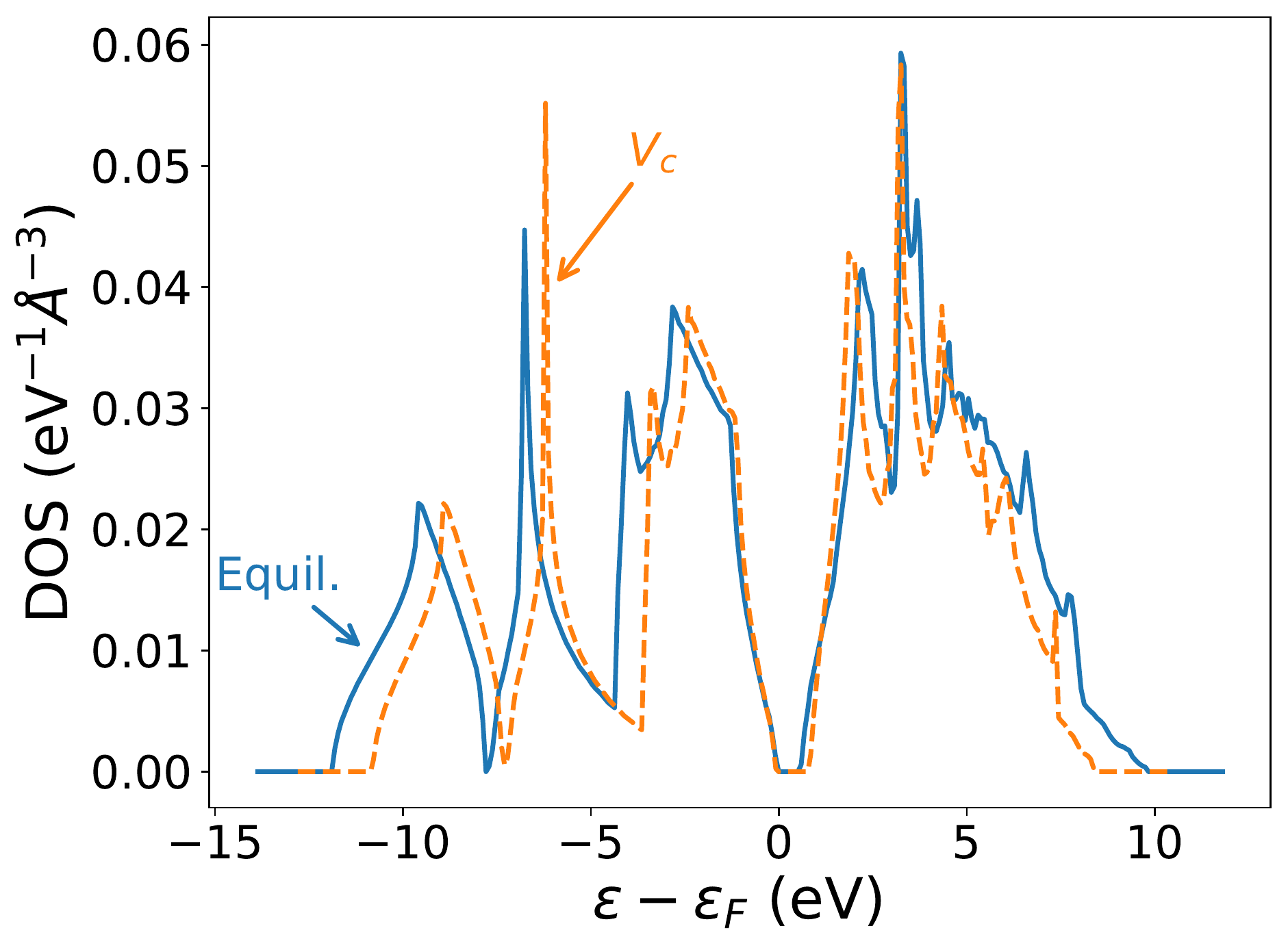}
  \caption{Intensive density of states plots for Si at both the PBE equilibrium lattice parameter 5.47 \AA{} (blue), and at the critical lattice parameter 5.81 \AA{} (orange). \label{fig:sidos}}
\end{figure}

\begin{figure}
  \includegraphics[width=\columnwidth]{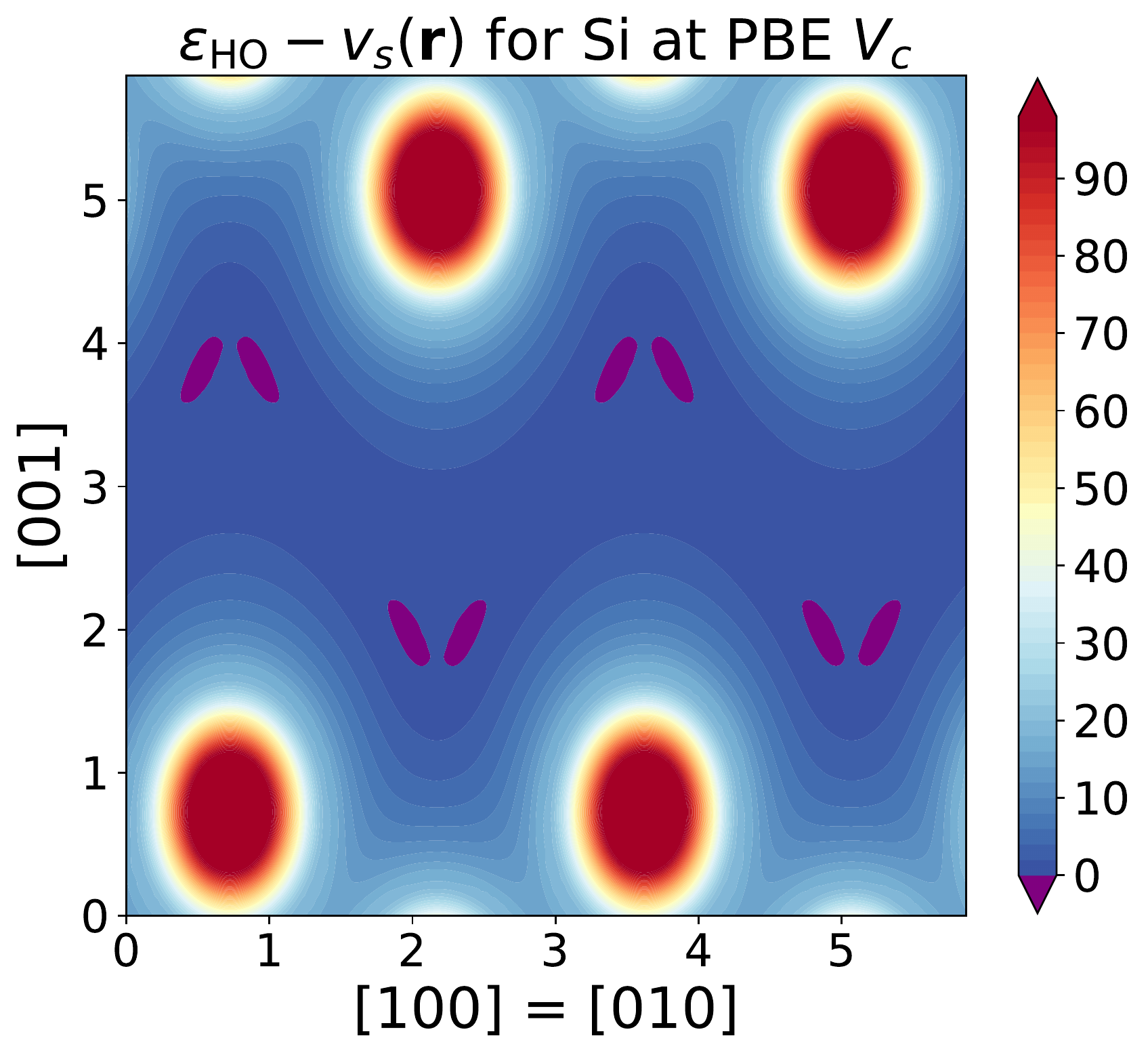}
  \caption{Contour plot of Si along the [110] (conventional cubic indices) direction at the PBE critical lattice constant $a_c = 5.79$ \AA{}, analogous to Fig. \ref{fig:ptmv} in the main text. The CFR (purple) is minute at this volume, about 0.53\% of the primitive cell volume, and located in the interstice. While this indicates the fit is not perfect, it provides a reasonable upper bound to $a_c$. \label{fig:sipot}}
\end{figure}

Here we include extra figures, fit parameters, and data tables that may prove useful for future work. A figure of the density of states for Si at its PBE equilibrium geometry, and at the critical lattice parameter, is given in Fig. \ref{fig:sidos}. A contour plot of $\homo - \vs(\br)$ in Si along the same plane as in Fig. \ref{fig:ptmv} of the main text is included in Fig. \ref{fig:sipot}.

\begin{figure}
  \centering
  \includegraphics[width=\columnwidth]{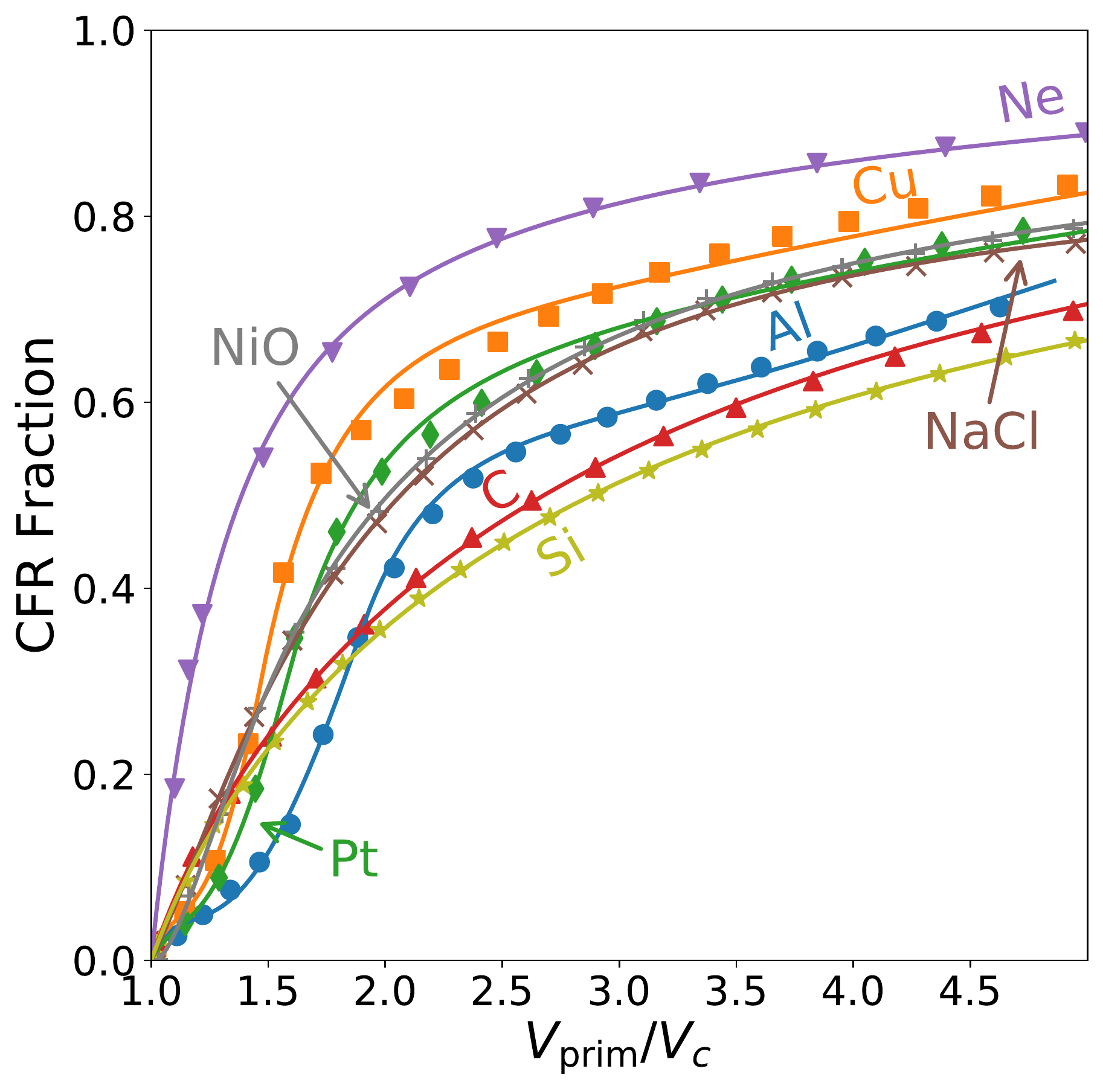}
  \caption{Figure analogous to Fig. \ref{fig:pbestrn} in the main text, but plotting $\vws/V_c$ to emphasize the shapes of the strain-CFR curves. Perhaps expectedly, the C and Si curves have exceedingly similar shapes. Quite unexpectedly, the NaCl and spin-unpolarized NiO curves are nearly identical. \label{fig:pbe_dimensionless}}
\end{figure}

\begin{figure}
  \centering
  \includegraphics[width=\columnwidth]{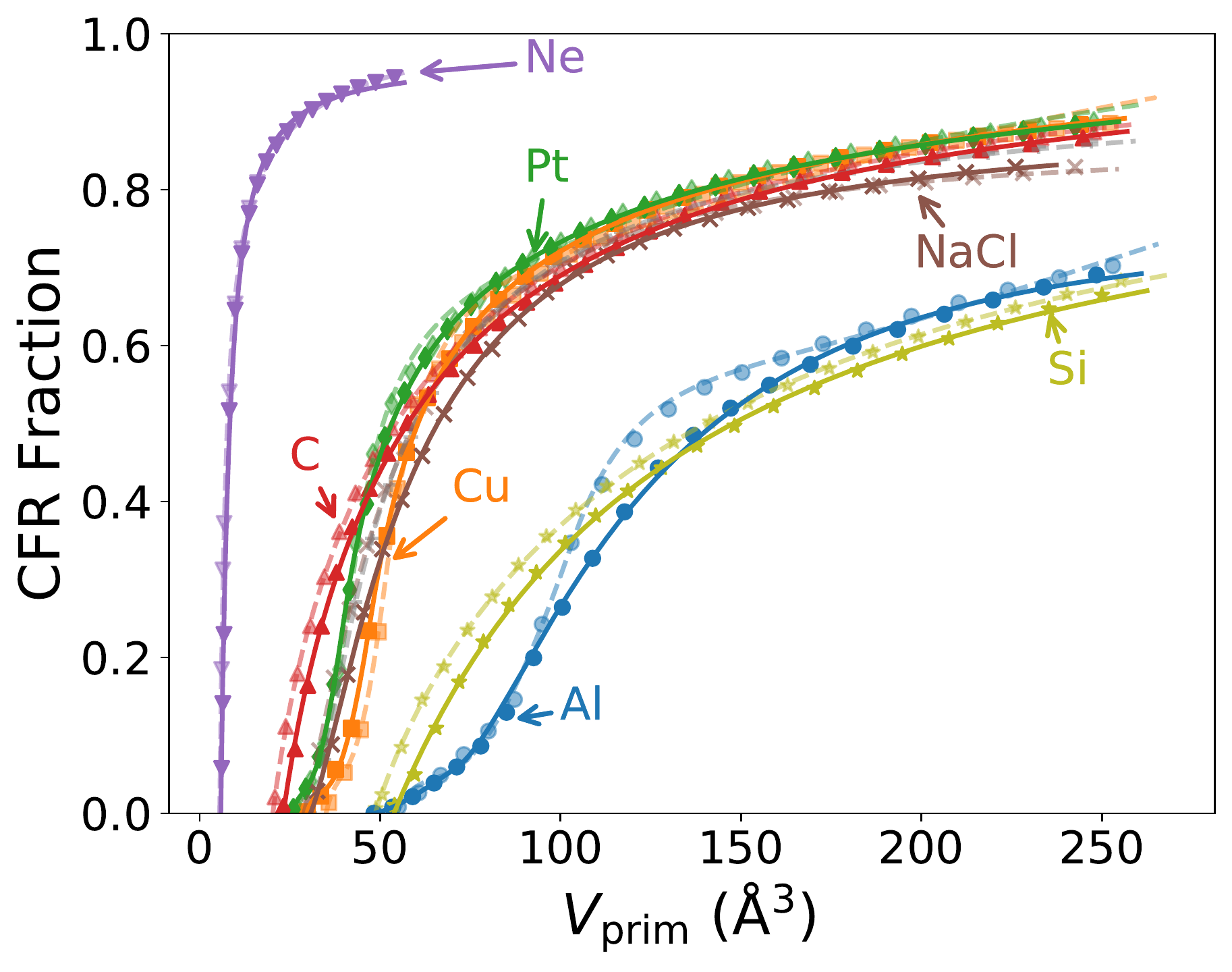}
  \caption{Emergence of LSDA (solid lines) and PBE (dashed faint lines) CFRs in Al (blue closed circles), Cu (yellow squares), Pt (green diamonds), C (red point-up triangles), Ne (purple point-down triangles), NaCl (brown crosses), and Si (olive stars) as a function of the lattice constant. All lines are fits given in Table \ref{tab:fits} in the main text. For the elemental insulators, the difference between the LSDA and PBE curves has the same sign. For the metals and NaCl, the LSDA and PBE curves cross.\label{fig:allstrn}}
\end{figure}

Fig. \ref{fig:pbe_dimensionless} is analogous to Fig. \ref{fig:pbestrn} of the main text, but emphasizes the shapes of the strain-CFR curves by plotting $\vws/V_c$. A plot of the LSDA fractional CFR volumes in the same manner as Fig. \ref{fig:pbestrn} in the main text, with PBE curves superposed faintly, is included in Fig. \ref{fig:allstrn}. Table \ref{tab:vac} enumerates fitted critical volumes $V_c$ for a comparison between VASP and Castep.

Last are a series of figures and tables for showing fitted strain curves for select main group elements as calculated with PBE in Castep, and using the fitting procedure in the main text, plotted separately for the Group 1, Group 2, Group 14, and Group 18 elemental solids.

We also include numerous tables (Tables \ref{tab:al_main_pbe_v_raw_dat}-\ref{tab:xe_group_18_pbe_c_raw_dat}) enumerating the raw data used for fitting and generating figures. All data presented is available in this text, and machine-readable data will be made available at reasonable request.

\clearpage

\clearpage

\begin{figure}[!ht]
  \centering
  \includegraphics[width=0.8\columnwidth]{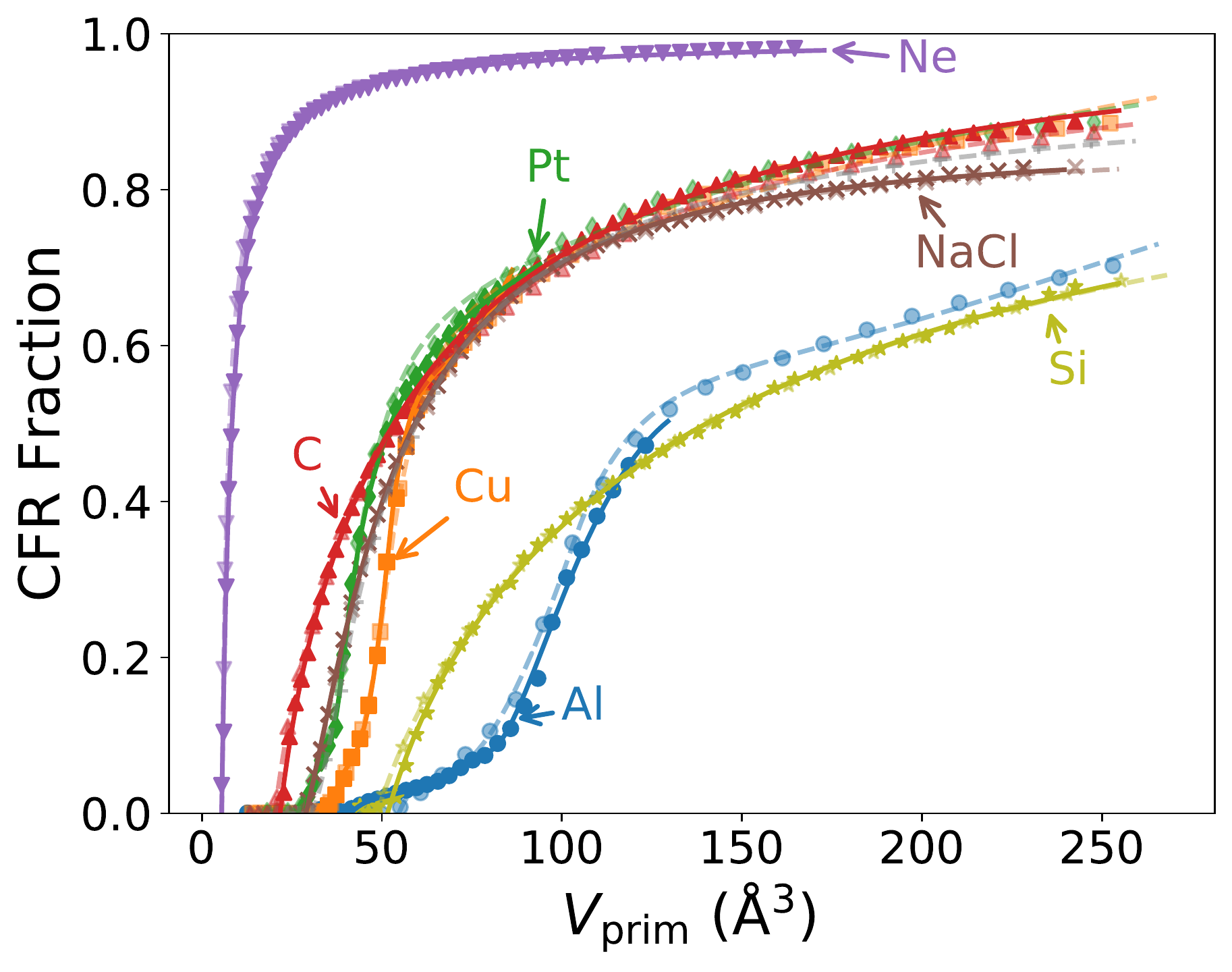}
  \caption{ Comparison between VASP and Castep results for PBE for the elements presented in the main text. \label{fig:pbe_comp}}
\end{figure}

\begin{table}[!h]
  \centering
  \begin{ruledtabular}
    \begin{tabular}{ccccccc}
      Solid (struc.) & $c_0$ & $c_1$ & $c_2$ & $c_3$ & $R^2$ & $V_c$ (\AA{}$^3$) \\ \hline
      \multirow{2}{*}{Al (fcc)} & 0.84 & -2.89 & 3.47 & -1.42 & 0.0010 & 41.33 \\
      & -2.69 & 27.97 & -76.55 & 63.47 & & 80.52 \\ \hline
      \multirow{2}{*}{Cu (fcc)} & 21.42 & -70.61 & 78.14 & -28.95 & 0.0011 & 36.43 \\
      & 6.71 & -32.96 & 59.87 & -36.80 & & 51.49 \\ \hline
      \multirow{2}{*}{Pt (fcc)} & 20.24 & -67.48 & 75.43 & -28.19 & 0.0008 & 28.89 \\
      & 1.22 & -2.94 & 5.52 & -4.64 & & 39.42 \\ \hline
      C (ds) & 1.04 & -1.73 & 1.18 & -0.50 & 0.0015 & 21.71 \\ \hline
      Ne (fcc) & 0.99 & -0.46 & -0.29 & -0.25 & 0.0061 & 5.49 \\ \hline
      NaCl (rs) & 0.86 & -0.09 & -1.85 & 1.09 & 0.0005 & 28.89 \\ \hline
      Si (ds) & 0.97 & -1.63 & 1.19 & -0.52 & 0.0014 & 51.54 \\
    \end{tabular}
    \caption{Fits for the solids presented in the main text as calculated with PBE in Castep. Materials with two lines of fit parameters use a separate fit for $\vws < V_0$ (first line) and $\vws \geq V_0$ (second) line. For these, $V_c$ is given on the first line, and $V_0$ is given on the second. For the fitting method and fit functions, refer to the main text. The fitted values of $V_c$ for Ne and Si are too large, and the fitted value of $V_c$ for Al is too small; all other values of $V_c$ are within their respective bounds from the numerical calculations. \label{tab:vac}}
  \end{ruledtabular}
\end{table}

\clearpage

\begin{figure}[!ht]
  \centering
  \begin{subfigure}
    \centering
    \includegraphics[width=0.49\columnwidth]{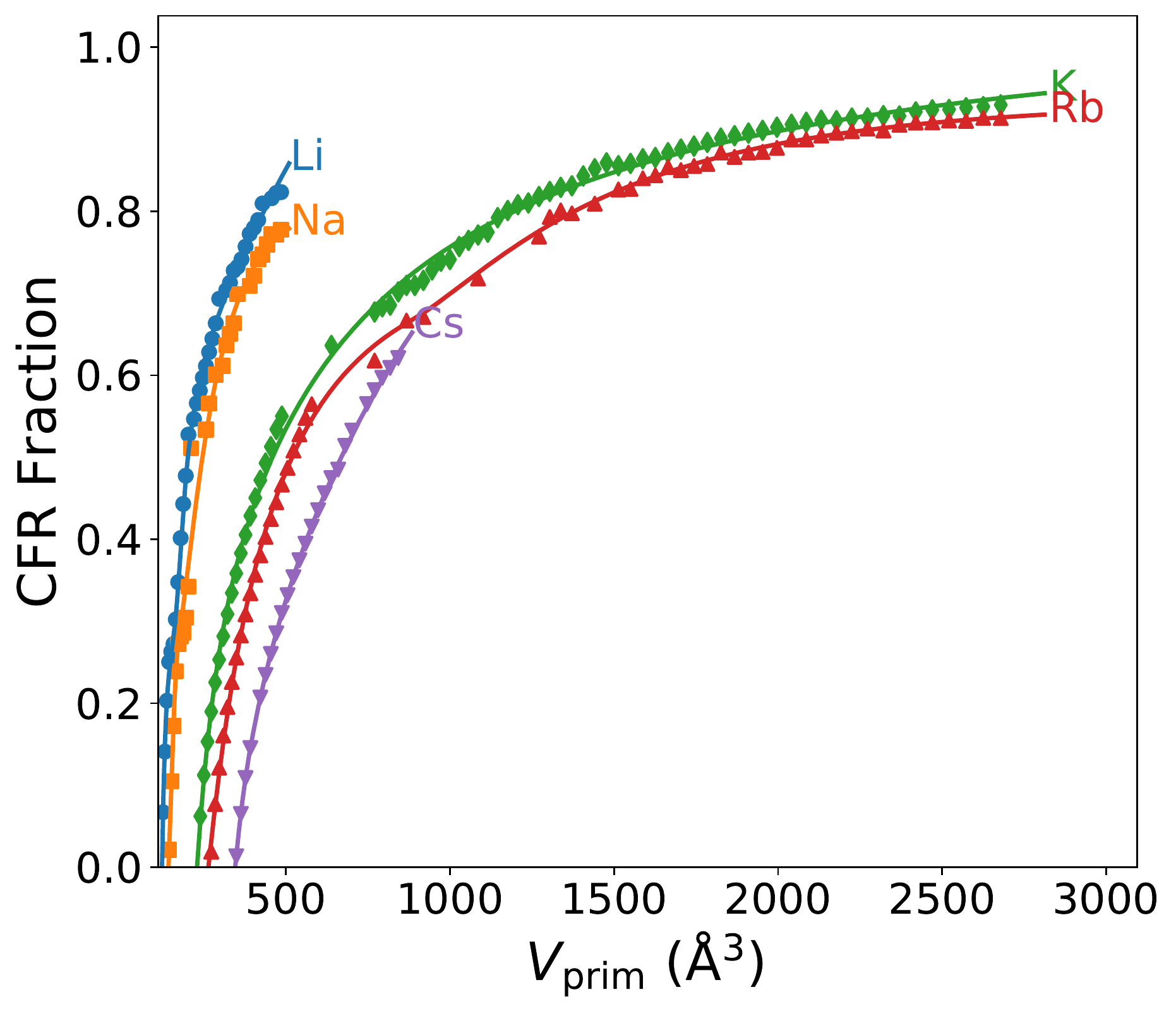}
    \label{fig:group1_vol}
  \end{subfigure}
  \begin{subfigure}
    \centering
    \includegraphics[width=0.49\columnwidth]{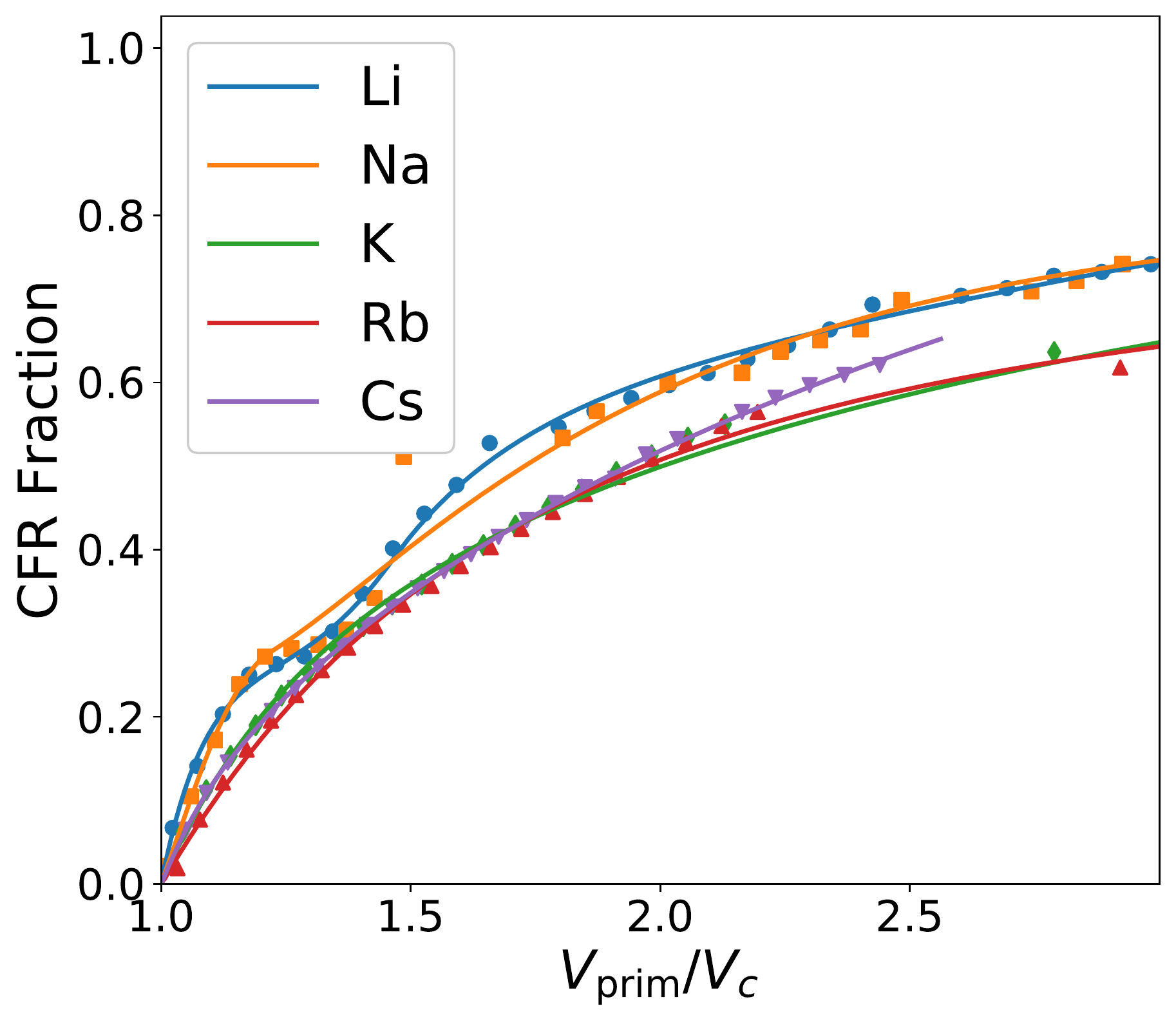}
    \label{fig:group1_red}
  \end{subfigure}
  \caption{ Fitted CFR strain curves, in dimensioned (left) and dimensionless (right) forms, for Group 1 (alkali metals) elemental solids as calculated with PBE in Castep. \label{fig:group1}}
\end{figure}

\begin{table}[!h]
  \centering
  \begin{ruledtabular}
    \begin{tabular}{cccccccc}
      Solid (struc.) & $c_0$ & $c_1$ & $c_2$ & $c_3$ & $R^2$ & $V_c$ (\AA{}$^3$) & $V_c/V_{\text{at}}$ \\ \hline
      \multirow{2}{*}{Li (bcc)} & 13.64 & -48.14 & 58.39 & -23.89 & 0.0032 & 122.20 & 1.50 \\
      & 1.51 & -4.29 & 7.92 & -5.93 & & 178.48 & \\ \hline
      \multirow{2}{*}{Na (bcc)} & -20.95 & 70.29 & -76.07 & 26.73 & 0.0180 & 141.92 & 1.37 \\
      & 0.66 & 1.52 & -4.77 & 2.85 & & 173.11 & \\ \hline
      K (bcc) & 1.07 & -1.67 & 1.50 & -0.90 & 0.0056 & 228.89 & 1.18 \\
      \multirow{2}{*}{Rb (bcc)} & 0.79 & -0.11 & -1.14 & 0.46 & 0.0023 & 263.60 & 1.11 \\
      & 0.89 & 1.54 & -15.46 & 26.25 & & 880.56 & \\ \hline
      \multirow{2}{*}{Cs (bcc)} & -1.43 & 5.39 & -5.04 & 1.08 & 0.0002 & 345.21 &  \\
      & 1.41 & -2.72 & 2.41 & -1.08 & & 391.33 & \\
    \end{tabular}
    \caption{ Fit parameters for the Group 1 (alkali metals) elemental solids as calculated with PBE in Castep. Materials with two lines of fit parameters use a separate fit for $\vws < V_0$ (first line) and $\vws \geq V_0$ (second) line. For these, $V_c$ is given on the first line, and $V_0$ is given on the second. For the fitting method and fit functions, refer to the main text. All values of $V_c$ are within their respective bounds from the numerical calculations. When possible, we report the ratio $V_c/V_{\text{at}}$, with $V_{\text{at}} = 4\pi r_{\text{TS}}^3/3$, a sphere at the non-relativistic turning surface radius $r_{\text{TS}}$ as reported in Ref. \cite{ospa2018}. \label{tab:group1}}
  \end{ruledtabular}
\end{table}

\clearpage

\begin{figure}[!ht]
  \centering
  \begin{subfigure}
    \centering
    \includegraphics[width=0.49\columnwidth]{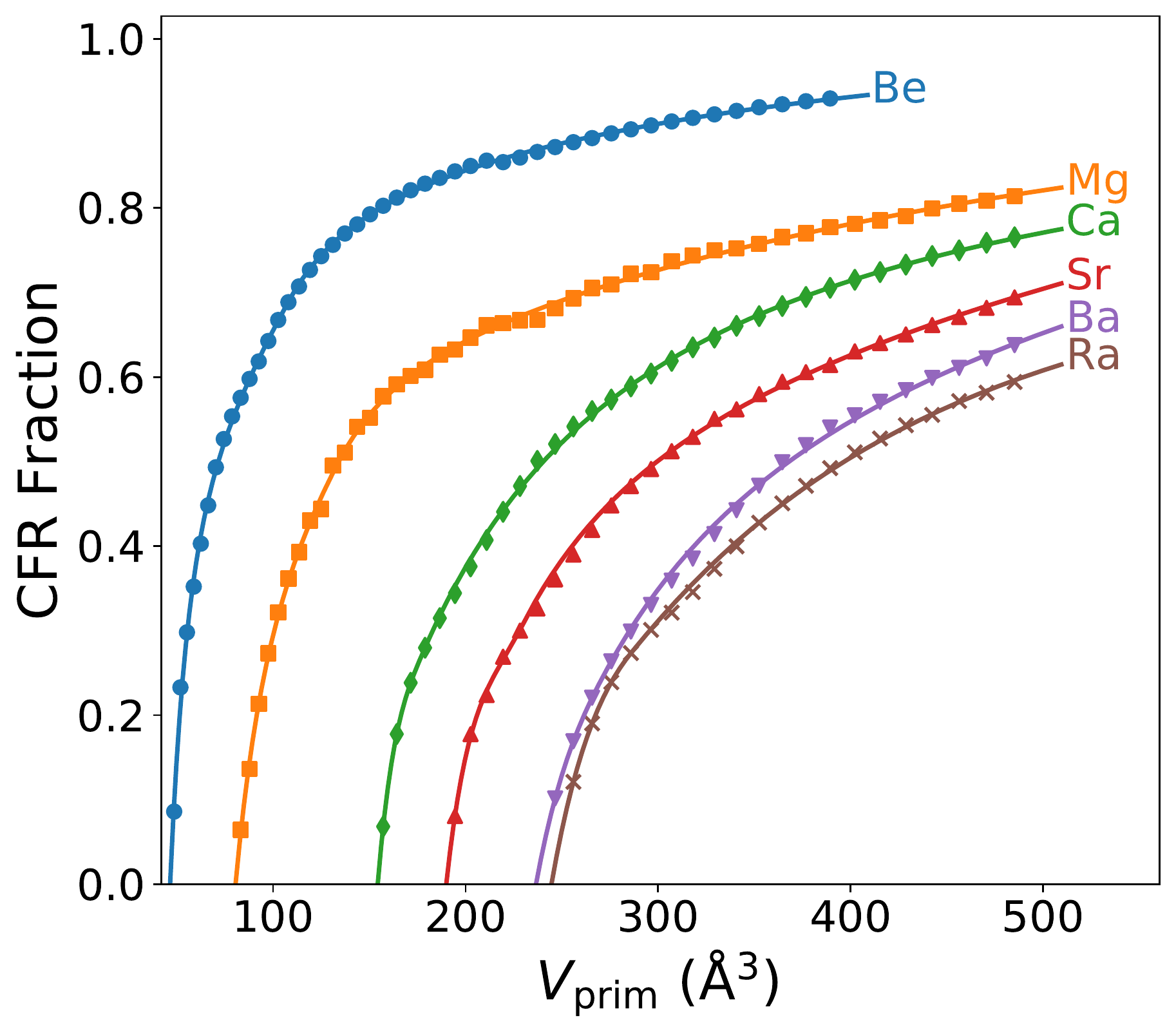}
    \label{fig:group2_vol}
  \end{subfigure}
  \begin{subfigure}
    \centering
    \includegraphics[width=0.49\columnwidth]{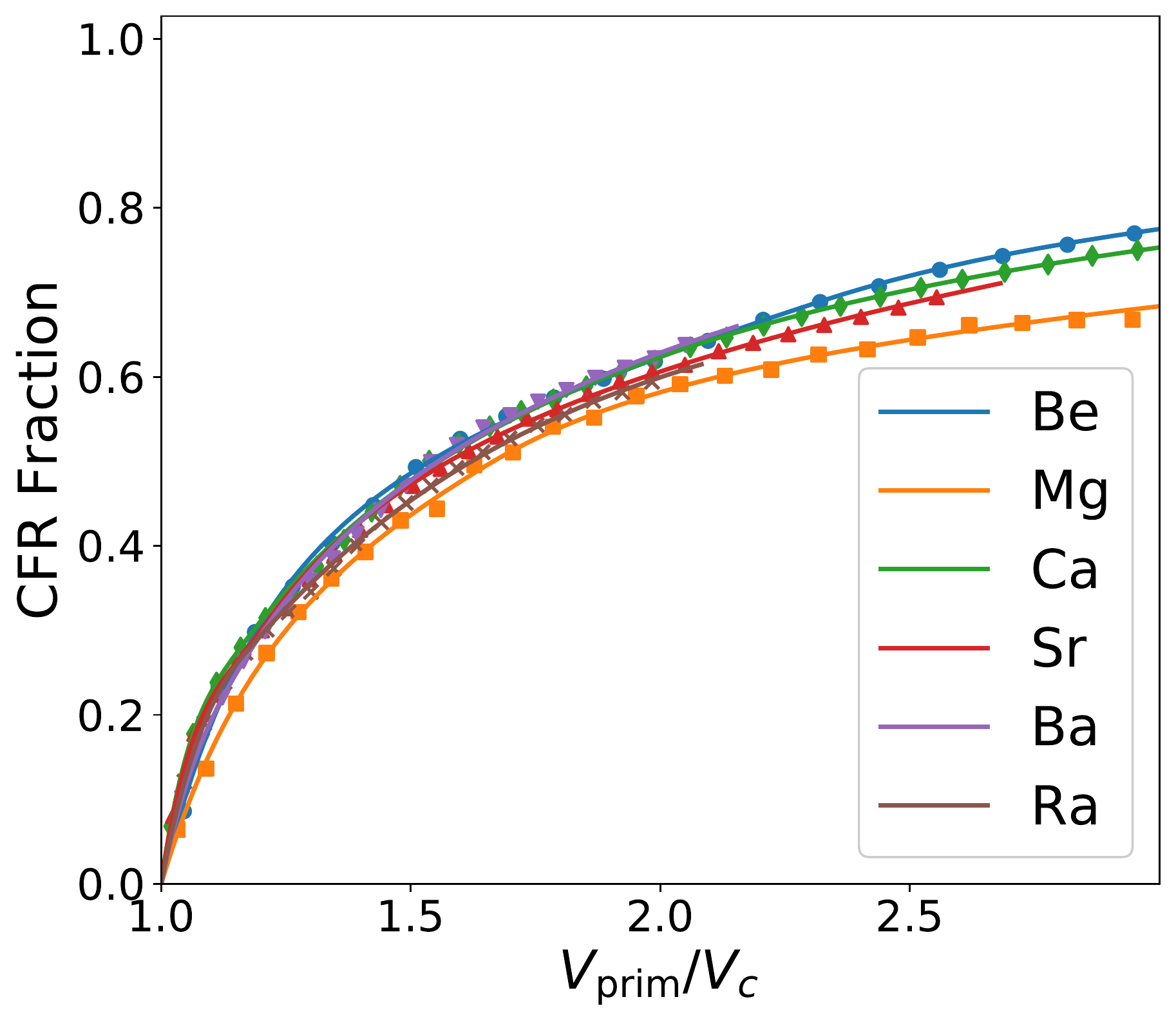}
    \label{fig:group2_red}
  \end{subfigure}
  \caption{ Fitted CFR strain curves, in dimensioned (left) and dimensionless (right) forms, for Group 2 (alkali earth metals) elemental solids as calculated with PBE in Castep. \label{fig:group2}}
\end{figure}

\begin{table}[!h]
  \centering
  \begin{ruledtabular}
    \begin{tabular}{cccccccc}
      Solid (struc.) & $c_0$ & $c_1$ & $c_2$ & $c_3$ & $R^2$ & $V_c$ (\AA{}$^3$) & $V_c/V_{\text{at}}$ \\ \hline
      \multirow{2}{*}{Be (bcc)} & 1.67 & -4.08 & 5.56 & -3.15 & 0.0011 & 46.55 & 1.94 \\
      & 1.06 & -1.39 & 2.89 & -3.84 & & 108.23 & \\ \hline
      \multirow{2}{*}{Mg (bcc)} & 1.86 & -4.58 & 5.56 & -2.84 & 0.0009 & 80.54 & 1.86 \\
      & 1.06 & -1.97 & 3.52 & -2.98 & & 133.80 & \\ \hline
      \multirow{2}{*}{Ca (fcc)} & 43.72 & -147.30 & 167.62 & -64.05 & 0.0004 & 154.37 & 1.66 \\
      & 0.97 & -0.59 & -0.17 & -0.09 & & 178.96 & \\ \hline
      \multirow{2}{*}{Sr (fcc)} & 38.23 & -129.32 & 147.96 & -56.87 & 0.0007 & 189.97 & 1.55 \\
      & 1.24 & -2.16 & 2.63 & -1.67 & & 229.06 & \\ \hline
      \multirow{2}{*}{Ba (bcc)} & 17.22 & -56.94 & 65.16 & -25.43 & 0.0007 & 236.59 &  \\
      & 1.20 & -1.68 & 1.62 & -1.09 & & 275.68 & \\ \hline
      \multirow{2}{*}{Ra (bcc)} & 8.39 & -30.21 & 38.44 & -16.62 & 0.0003 & 244.61 &  \\
      & 0.71 & 0.57 & -2.02 & 0.89 & & 285.89 & \\
    \end{tabular}
  \end{ruledtabular}
  \caption{ Fit parameters for the Group 2 (alkali earth metals) elemental solids as calculated with PBE in Castep. Materials with two lines of fit parameters use a separate fit for $\vws < V_0$ (first line) and $\vws \geq V_0$ (second) line. For these, $V_c$ is given on the first line, and $V_0$ is given on the second. For the fitting method and fit functions, refer to the main text. The fitted values of $V_c$ for Ba and Ra are too small; all other values of $V_c$ are within their respective bounds from the numerical calculations. When possible, we report the ratio $V_c/V_{\text{at}}$, with $V_{\text{at}} = 4\pi r_{\text{TS}}^3/3$, a sphere at the non-relativistic turning surface radius $r_{\text{TS}}$ as reported in Ref. \cite{ospa2018}. \label{tab:group2}}
\end{table}

\clearpage

\begin{figure}[!ht]
  \centering
  \begin{subfigure}
    \centering
    \includegraphics[width=0.49\columnwidth]{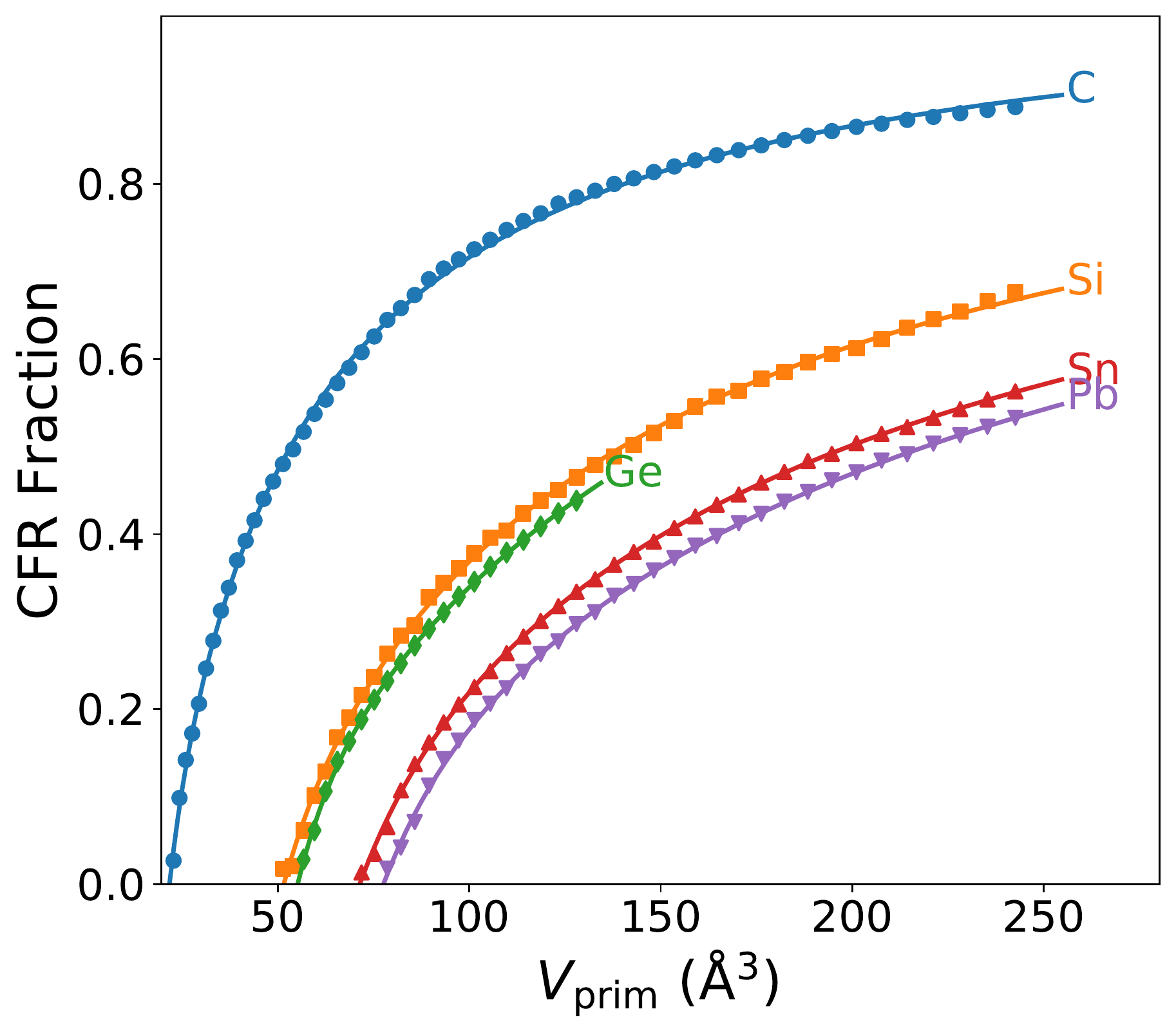}
    \label{fig:group14_vol}
  \end{subfigure}
  \begin{subfigure}
    \centering
    \includegraphics[width=0.49\columnwidth]{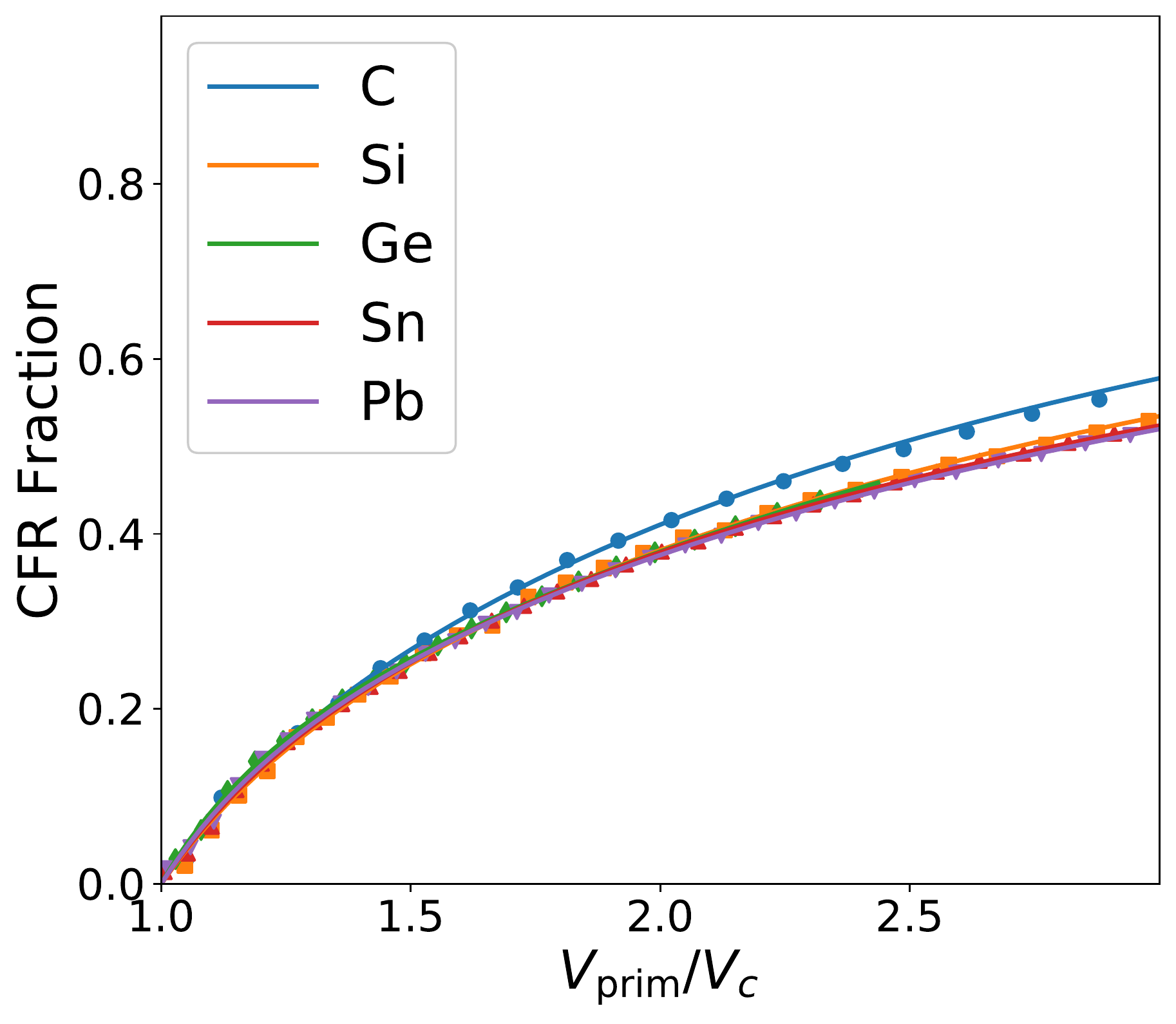}
    \label{fig:group14_red}
  \end{subfigure}
  \caption{ Fitted CFR strain curves, in dimensioned (left) and dimensionless (right) forms, for Group 14 (carbon group) elemental solids as calculated with PBE in Castep. \label{fig:group14}}
\end{figure}

\begin{table}[!h]
  \centering
  \begin{ruledtabular}
    \begin{tabular}{cccccccc}
      Solid (struc.) & $c_0$ & $c_1$ & $c_2$ & $c_3$ & $R^2$ & $V_c$ (\AA{}$^3$) & $V_c/V_{\text{at}}$ \\ \hline
      C (ds) & 1.04 & -1.73 & 1.18 & -0.50 & 0.0015 & 21.71 & 1.57 \\
      Si (ds) & 0.97 & -1.63 & 1.19 & -0.52 & 0.0014 & 51.54 & 1.24 \\
      Ge (ds) & 1.09 & -2.31 & 2.34 & -1.12 & 0.0001 & 55.17 & 1.22 \\
      Sn (ds) & 0.93 & -1.56 & 1.17 & -0.54 & 0.0003 & 71.37 & 1.17 \\
      Pb (ds) & 0.95 & -1.69 & 1.42 & -0.69 & 0.0003 & 77.56 &  \\
    \end{tabular}
  \caption{ Fit parameters for the Group 14 (carbon group) elemental solids as calculated with PBE in Castep. The fitted $V_c$ for Si is too large; all other values of $V_c$ are within their respective bounds from the numerical calculations. When possible, we report the ratio $V_c/V_{\text{at}}$, with $V_{\text{at}} = 4\pi r_{\text{TS}}^3/3$, a sphere at the non-relativistic turning surface radius $r_{\text{TS}}$ as reported in Ref. \cite{ospa2018}. \label{tab:group14}}
  \end{ruledtabular}
\end{table}

\clearpage

\begin{figure}[!ht]
  \centering
  \begin{subfigure}
    \centering
    \includegraphics[width=0.49\columnwidth]{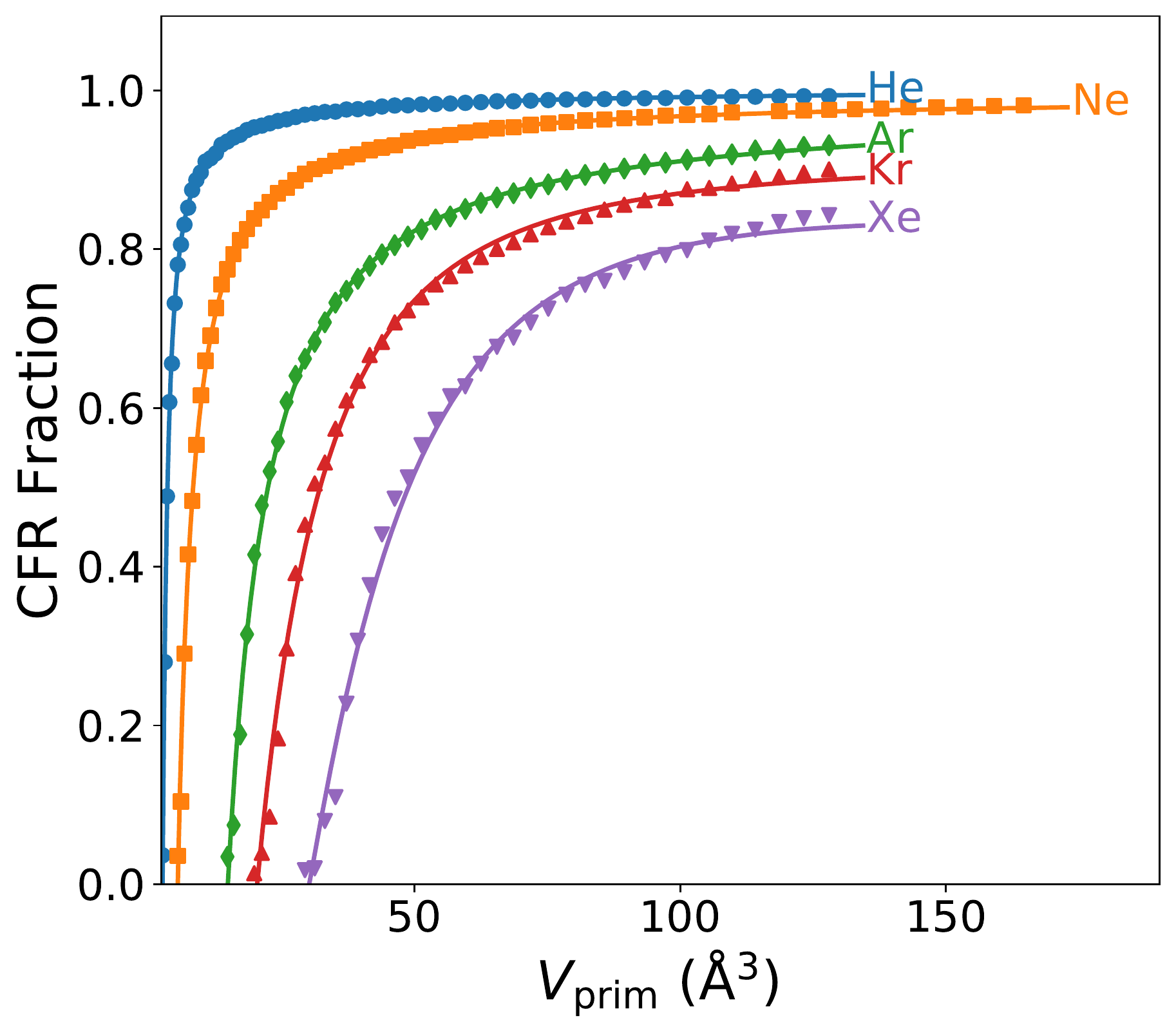}
    \label{fig:group18_vol}
  \end{subfigure}
  \begin{subfigure}
    \centering
    \includegraphics[width=0.49\columnwidth]{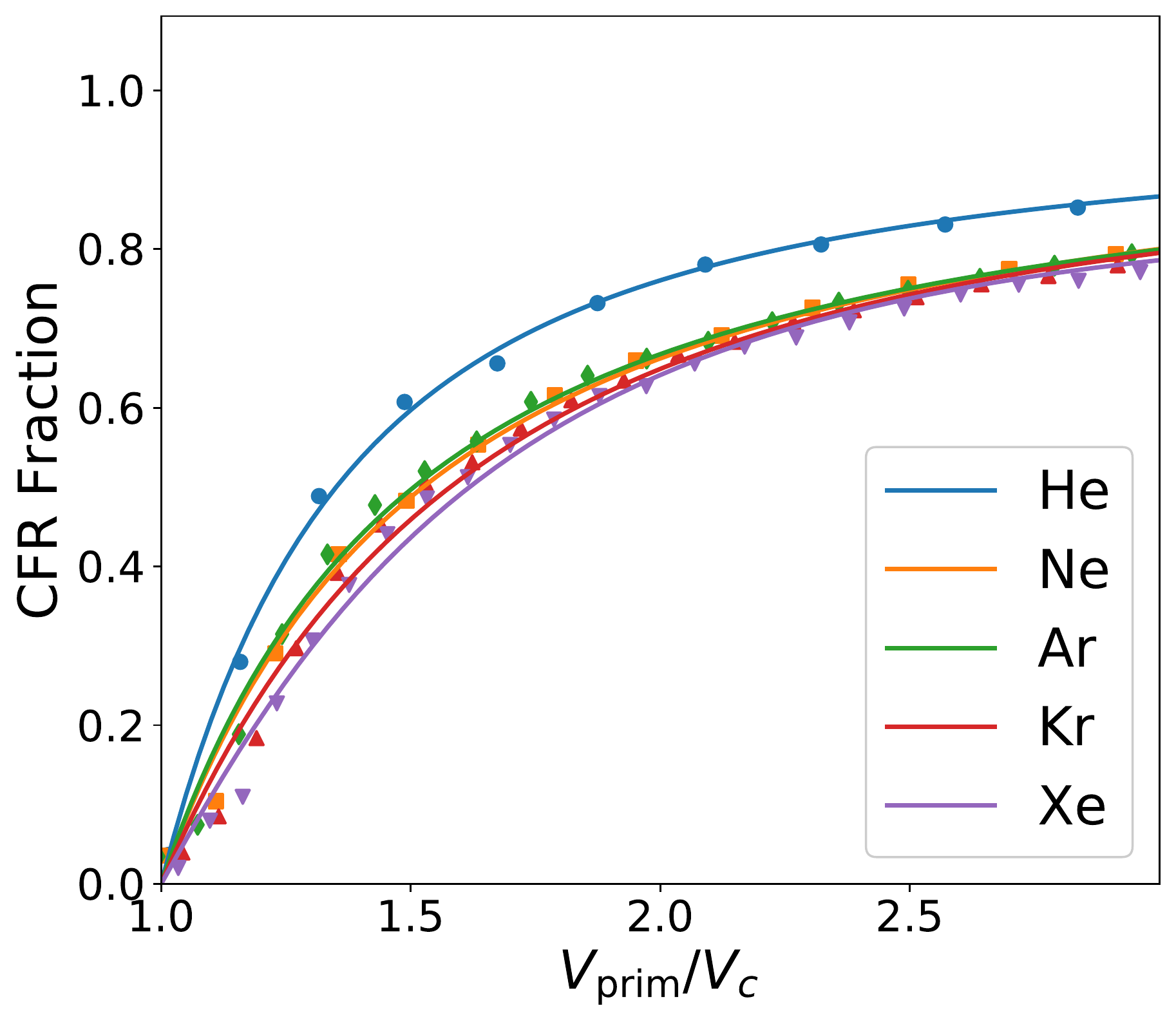}
    \label{fig:group18_red}
  \end{subfigure}
  \caption{ Fitted CFR strain curves, in dimensioned (left) and dimensionless (right) forms, for Group 18 (rare gases) elemental solids as calculated with PBE in Castep. \label{fig:group18}}
\end{figure}

\begin{table}[!h]
  \centering
  \begin{ruledtabular}
\begin{tabular}{cccccccc}
      Solid (struc.) & $c_0$ & $c_1$ & $c_2$ & $c_3$ & $R^2$ & $V_c$ (\AA{}$^3$) & $V_c/V_{\text{at}}$ \\ \hline
      He (fcc) & 1.00 & -0.40 & 0.27 & -0.87 & 0.0013 & 2.63 & 2.63 \\
      Ne (fcc) & 0.99 & -0.46 & -0.29 & -0.25 & 0.0061 & 5.49 & 1.99 \\
      Ar (fcc) & 0.98 & -0.45 & -0.19 & -0.34 & 0.0072 & 14.91 & 1.66 \\
      Kr (fcc) & 0.91 & 0.11 & -1.48 & 0.46 & 0.0139 & 20.43 & 1.54 \\
      Xe (fcc) & 0.74 & 1.03 & -3.15 & 1.38 & 0.0128 & 30.21 & 1.44 \\
    \end{tabular}
    \caption{ Fit parameters for the Group 18 (rare gas) elemental solids as calculated with PBE in Castep. All fitted values of $V_c$ are too large, except for He. When possible, we report the ratio $V_c/V_{\text{at}}$, with $V_{\text{at}} = 4\pi r_{\text{TS}}^3/3$, a sphere at the non-relativistic turning surface radius $r_{\text{TS}}$ as reported in Ref. \cite{ospa2018}. \label{tab:group18}}
  \end{ruledtabular}
\end{table}

\clearpage

\onecolumngrid
\section{Raw data for elements emphasized in main text}

\begin{longtable}{d{4}d{4}d{4}} 
  \caption{ Raw data for Al as calculated with PBE in VASP. \label{tab:al_main_pbe_v_raw_dat} } \\ \toprule 
       \multicolumn{1}{l}{$V_{\text{prim}}$ (\AA{}$^3$)} & \multicolumn{1}{l}{CFR fraction} & \multicolumn{1}{l}{Band gap from DOS (eV)} \\ \hline 
\endfirsthead 
\toprule
       \multicolumn{1}{l}{$V_{\text{prim}}$ (\AA{}$^3$)} & \multicolumn{1}{l}{CFR fraction} & \multicolumn{1}{l}{Band gap from DOS (eV)} \\ \hline 
\multicolumn{1}{c}{$\vdots$}& \multicolumn{1}{c}{$\vdots$}& \multicolumn{1}{c}{$\vdots$}\\ 
\endhead 
\multicolumn{1}{c}{$\vdots$}& \multicolumn{1}{c}{$\vdots$}& \multicolumn{1}{c}{$\vdots$}\\ 
\hline \hline 
\endfoot 
\hline \hline 
\endlastfoot
      16.4848 & 0.0000 & 0.0000 \\ 
      19.0563 & 0.0000 & 0.0000 \\ 
      21.8821 & 0.0000 & 0.0000 \\ 
      24.9743 & 0.0000 & 0.0000 \\ 
      28.3450 & 0.0000 & 0.0000 \\ 
      32.0060 & 0.0000 & 0.0000 \\ 
      35.9695 & 0.0000 & 0.0000 \\ 
      40.2473 & 0.0000 & 0.0000 \\ 
      44.8515 & 0.0000 & 0.0000 \\ 
      49.7942 & 0.0000 & 0.0000 \\ 
      55.0872 & 0.0081 & 0.0000 \\ 
      60.7427 & 0.0264 & 0.0000 \\ 
      66.7725 & 0.0491 & 0.0000 \\ 
      73.1887 & 0.0756 & 0.0000 \\ 
      80.0034 & 0.1056 & 0.0000 \\ 
      87.2284 & 0.1461 & 0.0000 \\ 
      94.8759 & 0.2427 & 0.0000 \\ 
      102.9577 & 0.3472 & 0.0000 \\ 
      111.4859 & 0.4219 & 0.0000 \\ 
      120.4726 & 0.4801 & 0.0000 \\ 
      129.9296 & 0.5185 & 0.0000 \\ 
      139.8691 & 0.5466 & 0.0000 \\ 
      150.3029 & 0.5656 & 0.0000 \\ 
      161.2431 & 0.5840 & 0.0000 \\ 
      172.7018 & 0.6022 & 0.0000 \\ 
      184.6908 & 0.6202 & 0.0000 \\ 
      197.2223 & 0.6378 & 0.0000 \\ 
      210.3081 & 0.6550 & 0.0000 \\ 
      223.9603 & 0.6712 & 0.0000 \\ 
      238.1910 & 0.6872 & 0.0000 \\ 
      253.0120 & 0.7024 & 0.0000 \\ 
\end{longtable} 
 
\clearpage

\begin{longtable}{d{4}d{4}d{4}} 
  \caption{ Raw data for Cu as calculated with PBE in VASP. \label{tab:cu_main_pbe_v_raw_dat} } \\ \toprule 
       \multicolumn{1}{l}{$V_{\text{prim}}$ (\AA{}$^3$)} & \multicolumn{1}{l}{CFR fraction} & \multicolumn{1}{l}{Band gap from DOS (eV)} \\ \hline 
\endfirsthead 
\toprule
       \multicolumn{1}{l}{$V_{\text{prim}}$ (\AA{}$^3$)} & \multicolumn{1}{l}{CFR fraction} & \multicolumn{1}{l}{Band gap from DOS (eV)} \\ \hline 
\multicolumn{1}{c}{$\vdots$}& \multicolumn{1}{c}{$\vdots$}& \multicolumn{1}{c}{$\vdots$}\\ 
\endhead 
\multicolumn{1}{c}{$\vdots$}& \multicolumn{1}{c}{$\vdots$}& \multicolumn{1}{c}{$\vdots$}\\ 
\hline \hline 
\endfoot 
\hline \hline 
\endlastfoot
      11.9580 & 0.0000 & 0.0000 \\ 
      14.0455 & 0.0000 & 0.0000 \\ 
      16.3627 & 0.0000 & 0.0000 \\ 
      18.9217 & 0.0000 & 0.0000 \\ 
      21.7346 & 0.0000 & 0.0000 \\ 
      24.8132 & 0.0000 & 0.0000 \\ 
      28.1696 & 0.0000 & 0.0000 \\ 
      31.8159 & 0.0000 & 0.0000 \\ 
      35.7639 & 0.0140 & 0.0000 \\ 
      40.0258 & 0.0523 & 0.0000 \\ 
      44.6134 & 0.1074 & 0.0000 \\ 
      49.5388 & 0.2328 & 0.0000 \\ 
      54.8141 & 0.4168 & 0.0000 \\ 
      60.4511 & 0.5237 & 0.0000 \\ 
      66.4619 & 0.5701 & 0.0000 \\ 
      72.8586 & 0.6038 & 0.0000 \\ 
      79.6530 & 0.6353 & 0.0000 \\ 
      86.8572 & 0.6651 & 0.0000 \\ 
      94.4833 & 0.6925 & 0.0000 \\ 
      102.5431 & 0.7171 & 0.0000 \\ 
      111.0487 & 0.7396 & 0.0000 \\ 
      120.0122 & 0.7599 & 0.0000 \\ 
      129.4454 & 0.7782 & 0.0000 \\ 
      139.3604 & 0.7944 & 0.0000 \\ 
      149.7693 & 0.8086 & 0.0000 \\ 
      160.6839 & 0.8219 & 0.0000 \\ 
      172.1163 & 0.8337 & 0.0000 \\ 
      184.0786 & 0.8446 & 0.0000 \\ 
      196.5826 & 0.8541 & 0.0000 \\ 
      209.6405 & 0.8632 & 0.0000 \\ 
      223.2641 & 0.8713 & 0.0000 \\ 
      237.4655 & 0.8785 & 0.0000 \\ 
      252.2568 & 0.8855 & 0.0000 \\ 
\end{longtable} 
 
\clearpage

\begin{longtable}{d{4}d{4}d{4}} 
  \caption{ Raw data for Pt as calculated with PBE in VASP. \label{tab:pt_main_pbe_v_raw_dat} } \\ \toprule 
       \multicolumn{1}{l}{$V_{\text{prim}}$ (\AA{}$^3$)} & \multicolumn{1}{l}{CFR fraction} & \multicolumn{1}{l}{Band gap from DOS (eV)} \\ \hline 
\endfirsthead 
\toprule
       \multicolumn{1}{l}{$V_{\text{prim}}$ (\AA{}$^3$)} & \multicolumn{1}{l}{CFR fraction} & \multicolumn{1}{l}{Band gap from DOS (eV)} \\ \hline 
\multicolumn{1}{c}{$\vdots$}& \multicolumn{1}{c}{$\vdots$}& \multicolumn{1}{c}{$\vdots$}\\ 
\endhead 
\multicolumn{1}{c}{$\vdots$}& \multicolumn{1}{c}{$\vdots$}& \multicolumn{1}{c}{$\vdots$}\\ 
\hline \hline 
\endfoot 
\hline \hline 
\endlastfoot
      15.6427 & 0.0000 & 0.0000 \\ 
      18.1279 & 0.0000 & 0.0000 \\ 
      20.8634 & 0.0000 & 0.0000 \\ 
      23.8610 & 0.0000 & 0.0000 \\ 
      27.1328 & 0.0059 & 0.0000 \\ 
      30.6909 & 0.0411 & 0.0000 \\ 
      34.5471 & 0.0890 & 0.0000 \\ 
      38.7135 & 0.1846 & 0.0000 \\ 
      43.2022 & 0.3464 & 0.0000 \\ 
      48.0250 & 0.4608 & 0.0000 \\ 
      53.1940 & 0.5255 & 0.0000 \\ 
      58.7213 & 0.5652 & 0.0000 \\ 
      64.6187 & 0.5994 & 0.0000 \\ 
      70.8983 & 0.6311 & 0.0000 \\ 
      77.5722 & 0.6605 & 0.0000 \\ 
      84.6522 & 0.6871 & 0.0000 \\ 
      92.1505 & 0.7105 & 0.0000 \\ 
      100.0789 & 0.7317 & 0.0000 \\ 
      108.4495 & 0.7510 & 0.0000 \\ 
      117.2744 & 0.7687 & 0.0000 \\ 
      126.5654 & 0.7844 & 0.0000 \\ 
      136.3346 & 0.7990 & 0.0000 \\ 
      146.5941 & 0.8121 & 0.0000 \\ 
      157.3557 & 0.8244 & 0.0000 \\ 
      168.6315 & 0.8355 & 0.0000 \\ 
      180.4336 & 0.8457 & 0.0000 \\ 
      192.7738 & 0.8550 & 0.0000 \\ 
      205.6642 & 0.8637 & 0.0000 \\ 
      219.1169 & 0.8718 & 0.0000 \\ 
      233.1437 & 0.8793 & 0.0000 \\ 
      247.7567 & 0.8863 & 0.0000 \\ 
\end{longtable} 
 
\clearpage

\begin{longtable}{d{4}d{4}d{4}} 
  \caption{ Raw data for C as calculated with PBE in VASP. \label{tab:c_main_pbe_v_raw_dat} } \\ \toprule 
       \multicolumn{1}{l}{$V_{\text{prim}}$ (\AA{}$^3$)} & \multicolumn{1}{l}{CFR fraction} & \multicolumn{1}{l}{Band gap from DOS (eV)} \\ \hline 
\endfirsthead 
\toprule
       \multicolumn{1}{l}{$V_{\text{prim}}$ (\AA{}$^3$)} & \multicolumn{1}{l}{CFR fraction} & \multicolumn{1}{l}{Band gap from DOS (eV)} \\ \hline 
\multicolumn{1}{c}{$\vdots$}& \multicolumn{1}{c}{$\vdots$}& \multicolumn{1}{c}{$\vdots$}\\ 
\endhead 
\multicolumn{1}{c}{$\vdots$}& \multicolumn{1}{c}{$\vdots$}& \multicolumn{1}{c}{$\vdots$}\\ 
\hline \hline 
\endfoot 
\hline \hline 
\endlastfoot
      11.3748 & 0.0000 & 4.1395 \\ 
      13.3957 & 0.0000 & 3.7061 \\ 
      15.6427 & 0.0000 & 3.4592 \\ 
      18.1279 & 0.0000 & 3.0474 \\ 
      20.8634 & 0.0212 & 2.3060 \\ 
      23.8610 & 0.1116 & 0.5534 \\ 
      27.1328 & 0.1794 & 0.0000 \\ 
      30.6909 & 0.2407 & 0.0000 \\ 
      34.5471 & 0.3036 & 0.0000 \\ 
      38.7135 & 0.3615 & 0.0000 \\ 
      43.2022 & 0.4113 & 0.0000 \\ 
      48.0250 & 0.4548 & 0.0000 \\ 
      53.1940 & 0.4946 & 0.0000 \\ 
      58.7213 & 0.5302 & 0.0000 \\ 
      64.6187 & 0.5637 & 0.0000 \\ 
      70.8983 & 0.5943 & 0.0000 \\ 
      77.5722 & 0.6228 & 0.0000 \\ 
      84.6522 & 0.6492 & 0.0000 \\ 
      92.1505 & 0.6747 & 0.0000 \\ 
      100.0789 & 0.6986 & 0.0000 \\ 
      108.4495 & 0.7214 & 0.0000 \\ 
      117.2744 & 0.7436 & 0.0000 \\ 
      126.5654 & 0.7654 & 0.0000 \\ 
      136.3346 & 0.7839 & 0.0000 \\ 
      146.5941 & 0.7981 & 0.0000 \\ 
      157.3557 & 0.8104 & 0.0000 \\ 
      168.6315 & 0.8217 & 0.0000 \\ 
      180.4336 & 0.8321 & 0.0000 \\ 
      192.7738 & 0.8421 & 0.0000 \\ 
      205.6642 & 0.8509 & 0.0000 \\ 
      219.1169 & 0.8593 & 0.0000 \\ 
      233.1437 & 0.8669 & 0.0000 \\ 
      247.7567 & 0.8743 & 0.0000 \\ 
\end{longtable} 
 
\clearpage

\begin{longtable}{d{4}d{4}d{4}} 
  \caption{ Raw data for Ne as calculated with PBE in VASP. \label{tab:ne_main_pbe_v_raw_dat} } \\ \toprule 
       \multicolumn{1}{l}{$V_{\text{prim}}$ (\AA{}$^3$)} & \multicolumn{1}{l}{CFR fraction} & \multicolumn{1}{l}{Band gap from DOS (eV)} \\ \hline 
\endfirsthead 
\toprule
       \multicolumn{1}{l}{$V_{\text{prim}}$ (\AA{}$^3$)} & \multicolumn{1}{l}{CFR fraction} & \multicolumn{1}{l}{Band gap from DOS (eV)} \\ \hline 
\multicolumn{1}{c}{$\vdots$}& \multicolumn{1}{c}{$\vdots$}& \multicolumn{1}{c}{$\vdots$}\\ 
\endhead 
\multicolumn{1}{c}{$\vdots$}& \multicolumn{1}{c}{$\vdots$}& \multicolumn{1}{c}{$\vdots$}\\ 
\hline \hline 
\endfoot 
\hline \hline 
\endlastfoot
      5.7873 & 0.0933 & 18.5758 \\ 
      6.0972 & 0.1847 & 17.9035 \\ 
      6.4181 & 0.3122 & 17.5081 \\ 
      6.7500 & 0.3719 & 16.9491 \\ 
      8.1920 & 0.5404 & 14.9407 \\ 
      9.8260 & 0.6535 & 13.8288 \\ 
      11.6640 & 0.7240 & 12.9718 \\ 
      13.7180 & 0.7763 & 12.4658 \\ 
      16.0000 & 0.8088 & 12.1625 \\ 
      18.5220 & 0.8356 & 11.7600 \\ 
      21.2960 & 0.8570 & 11.5588 \\ 
      24.3340 & 0.8745 & 11.4974 \\ 
      27.6480 & 0.8899 & 11.2233 \\ 
      31.2500 & 0.9024 & 11.2782 \\ 
      35.1520 & 0.9131 & 11.2431 \\ 
      39.3660 & 0.9227 & 11.3382 \\ 
      43.9040 & 0.9306 & 11.3177 \\ 
      48.7780 & 0.9373 & 11.3249 \\ 
      54.0000 & 0.9436 & 11.3459 \\ 
\end{longtable} 
 
\clearpage

\begin{longtable}{d{4}d{4}d{4}} 
  \caption{ Raw data for NaCl as calculated with PBE in VASP. \label{tab:nacl_main_pbe_v_raw_dat} } \\ \toprule 
       \multicolumn{1}{l}{$V_{\text{prim}}$ (\AA{}$^3$)} & \multicolumn{1}{l}{CFR fraction} & \multicolumn{1}{l}{Band gap from DOS (eV)} \\ \hline 
\endfirsthead 
\toprule
       \multicolumn{1}{l}{$V_{\text{prim}}$ (\AA{}$^3$)} & \multicolumn{1}{l}{CFR fraction} & \multicolumn{1}{l}{Band gap from DOS (eV)} \\ \hline 
\multicolumn{1}{c}{$\vdots$}& \multicolumn{1}{c}{$\vdots$}& \multicolumn{1}{c}{$\vdots$}\\ 
\endhead 
\multicolumn{1}{c}{$\vdots$}& \multicolumn{1}{c}{$\vdots$}& \multicolumn{1}{c}{$\vdots$}\\ 
\hline \hline 
\endfoot 
\hline \hline 
\endlastfoot
      12.6633 & 0.0000 & 0.2540 \\ 
      14.8297 & 0.0000 & 2.3167 \\ 
      17.2302 & 0.0000 & 3.8483 \\ 
      19.8767 & 0.0000 & 4.9494 \\ 
      22.7813 & 0.0000 & 5.9849 \\ 
      25.9558 & 0.0000 & 6.6959 \\ 
      29.4123 & 0.0134 & 7.1671 \\ 
      33.1627 & 0.0809 & 6.5604 \\ 
      37.2193 & 0.1739 & 5.9844 \\ 
      41.5938 & 0.2617 & 5.3491 \\ 
      46.2983 & 0.3438 & 4.9802 \\ 
      51.3447 & 0.4150 & 4.5501 \\ 
      56.7452 & 0.4699 & 4.3312 \\ 
      62.5117 & 0.5212 & 4.0477 \\ 
      68.6562 & 0.5700 & 3.7484 \\ 
      75.1907 & 0.6093 & 3.6438 \\ 
      82.1273 & 0.6404 & 3.3609 \\ 
      89.4778 & 0.6765 & 3.3206 \\ 
      97.2542 & 0.6987 & 3.0762 \\ 
      105.4688 & 0.7181 & 3.0424 \\ 
      114.1333 & 0.7342 & 2.9033 \\ 
      123.2598 & 0.7463 & 2.6795 \\ 
      132.8603 & 0.7611 & 2.6415 \\ 
      142.9468 & 0.7705 & 2.5432 \\ 
      153.5312 & 0.7796 & 2.4179 \\ 
      164.6258 & 0.7896 & 2.2904 \\ 
      176.2422 & 0.7971 & 2.1310 \\ 
      188.3928 & 0.8059 & 2.1254 \\ 
      201.0893 & 0.8094 & 1.9031 \\ 
      214.3438 & 0.8160 & 1.7761 \\ 
      228.1683 & 0.8215 & 1.6488 \\ 
      242.5748 & 0.8291 & 1.6157 \\ 
\end{longtable} 
 
\clearpage

\begin{longtable}{d{4}d{4}d{4}} 
  \caption{ Raw data for Si as calculated with PBE in VASP. \label{tab:si_main_pbe_v_raw_dat} } \\ \toprule 
       \multicolumn{1}{l}{$V_{\text{prim}}$ (\AA{}$^3$)} & \multicolumn{1}{l}{CFR fraction} & \multicolumn{1}{l}{Band gap from DOS (eV)} \\ \hline 
\endfirsthead 
\toprule
       \multicolumn{1}{l}{$V_{\text{prim}}$ (\AA{}$^3$)} & \multicolumn{1}{l}{CFR fraction} & \multicolumn{1}{l}{Band gap from DOS (eV)} \\ \hline 
\multicolumn{1}{c}{$\vdots$}& \multicolumn{1}{c}{$\vdots$}& \multicolumn{1}{c}{$\vdots$}\\ 
\endhead 
\multicolumn{1}{c}{$\vdots$}& \multicolumn{1}{c}{$\vdots$}& \multicolumn{1}{c}{$\vdots$}\\ 
\hline \hline 
\endfoot 
\hline \hline 
\endlastfoot
      40.9168 & 0.0000 & 0.5507 \\ 
      45.5711 & 0.0000 & 0.7547 \\ 
      48.7857 & 0.0064 & 0.8074 \\ 
      50.5655 & 0.0242 & 0.7107 \\ 
      55.9121 & 0.0849 & 0.0000 \\ 
      61.6230 & 0.1457 & 0.0000 \\ 
      67.7100 & 0.1885 & 0.0000 \\ 
      74.1852 & 0.2351 & 0.0000 \\ 
      81.0607 & 0.2779 & 0.0000 \\ 
      88.3483 & 0.3186 & 0.0000 \\ 
      96.0601 & 0.3554 & 0.0000 \\ 
      104.2082 & 0.3890 & 0.0000 \\ 
      112.8044 & 0.4199 & 0.0000 \\ 
      121.8609 & 0.4493 & 0.0000 \\ 
      131.3895 & 0.4764 & 0.0000 \\ 
      141.4023 & 0.5022 & 0.0000 \\ 
      151.9114 & 0.5264 & 0.0000 \\ 
      162.9286 & 0.5491 & 0.0000 \\ 
      174.4660 & 0.5709 & 0.0000 \\ 
      186.5357 & 0.5916 & 0.0000 \\ 
      199.1495 & 0.6114 & 0.0000 \\ 
      212.3195 & 0.6304 & 0.0000 \\ 
      226.0578 & 0.6488 & 0.0000 \\ 
      240.3762 & 0.6661 & 0.0000 \\ 
      255.2868 & 0.6834 & 0.0000 \\ 
\end{longtable} 
 
\clearpage

\begin{longtable}{d{4}d{4}d{4}} 
  \caption{ Raw data for spin-unpolarized NiO as calculated with PBE in VASP. \label{tab:nio_main_pbe_v_raw_dat} } \\ \toprule 
       \multicolumn{1}{l}{$V_{\text{prim}}$ (\AA{}$^3$)} & \multicolumn{1}{l}{CFR fraction} & \multicolumn{1}{l}{Band gap from DOS (eV)} \\ \hline 
\endfirsthead 
\toprule
       \multicolumn{1}{l}{$V_{\text{prim}}$ (\AA{}$^3$)} & \multicolumn{1}{l}{CFR fraction} & \multicolumn{1}{l}{Band gap from DOS (eV)} \\ \hline 
\multicolumn{1}{c}{$\vdots$}& \multicolumn{1}{c}{$\vdots$}& \multicolumn{1}{c}{$\vdots$}\\ 
\endhead 
\multicolumn{1}{c}{$\vdots$}& \multicolumn{1}{c}{$\vdots$}& \multicolumn{1}{c}{$\vdots$}\\ 
\hline \hline 
\endfoot 
\hline \hline 
\endlastfoot
      17.9978 & 0.0000 & 0.0000 \\ 
      20.7205 & 0.0000 & 0.0000 \\ 
      23.7047 & 0.0000 & 0.0000 \\ 
      26.9625 & 0.0000 & 0.0000 \\ 
      30.5060 & 0.0007 & 0.0000 \\ 
      34.3470 & 0.0691 & 0.0000 \\ 
      38.4977 & 0.1572 & 0.0000 \\ 
      42.9699 & 0.2711 & 0.0000 \\ 
      47.7757 & 0.3531 & 0.0000 \\ 
      52.9272 & 0.4213 & 0.0000 \\ 
      58.4362 & 0.4829 & 0.0000 \\ 
      64.3149 & 0.5394 & 0.0000 \\ 
      70.5751 & 0.5882 & 0.0000 \\ 
      77.2289 & 0.6259 & 0.0000 \\ 
      84.2884 & 0.6593 & 0.0000 \\ 
      91.7654 & 0.6881 & 0.0000 \\ 
      99.6721 & 0.7116 & 0.0000 \\ 
      108.0203 & 0.7293 & 0.0000 \\ 
      116.8221 & 0.7454 & 0.0000 \\ 
      126.0896 & 0.7604 & 0.2030 \\ 
      135.8346 & 0.7739 & 0.0000 \\ 
      146.0693 & 0.7866 & 0.0000 \\ 
      156.8055 & 0.7986 & 0.2018 \\ 
      168.0553 & 0.8097 & 0.1669 \\ 
      179.8308 & 0.8203 & 0.0000 \\ 
      192.1438 & 0.8300 & 0.0000 \\ 
      205.0065 & 0.8389 & 0.1942 \\ 
      218.4307 & 0.8476 & 0.1132 \\ 
      232.4285 & 0.8553 & 0.1173 \\ 
      247.0120 & 0.8628 & 0.0000 \\ 
\end{longtable} 
 
\clearpage

\begin{longtable}{d{4}d{4}d{4}} 
  \caption{ Raw data for Al as calculated with LSDA in VASP. \label{tab:al_main_lsd_v_raw_dat} } \\ \toprule 
       \multicolumn{1}{l}{$V_{\text{prim}}$ (\AA{}$^3$)} & \multicolumn{1}{l}{CFR fraction} & \multicolumn{1}{l}{Band gap from DOS (eV)} \\ \hline 
\endfirsthead 
\toprule
       \multicolumn{1}{l}{$V_{\text{prim}}$ (\AA{}$^3$)} & \multicolumn{1}{l}{CFR fraction} & \multicolumn{1}{l}{Band gap from DOS (eV)} \\ \hline 
\multicolumn{1}{c}{$\vdots$}& \multicolumn{1}{c}{$\vdots$}& \multicolumn{1}{c}{$\vdots$}\\ 
\endhead 
\multicolumn{1}{c}{$\vdots$}& \multicolumn{1}{c}{$\vdots$}& \multicolumn{1}{c}{$\vdots$}\\ 
\hline \hline 
\endfoot 
\hline \hline 
\endlastfoot
      15.7612 & 0.0000 & 0.0000 \\ 
      18.2587 & 0.0000 & 0.0000 \\ 
      21.0069 & 0.0000 & 0.0000 \\ 
      24.0180 & 0.0000 & 0.0000 \\ 
      27.3038 & 0.0000 & 0.0000 \\ 
      30.8765 & 0.0000 & 0.0000 \\ 
      34.7480 & 0.0000 & 0.0000 \\ 
      38.9302 & 0.0000 & 0.0000 \\ 
      43.4353 & 0.0000 & 0.0000 \\ 
      48.2751 & 0.0002 & 0.0000 \\ 
      53.4618 & 0.0081 & 0.0000 \\ 
      59.0073 & 0.0215 & 0.0000 \\ 
      64.9235 & 0.0390 & 0.0000 \\ 
      71.2226 & 0.0597 & 0.0000 \\ 
      77.9164 & 0.0865 & 0.0000 \\ 
      85.0171 & 0.1297 & 0.0000 \\ 
      92.5366 & 0.1997 & 0.0000 \\ 
      100.4868 & 0.2644 & 0.0000 \\ 
      108.8799 & 0.3272 & 0.0000 \\ 
      117.7277 & 0.3868 & 0.0000 \\ 
      127.0424 & 0.4432 & 0.0000 \\ 
      136.8359 & 0.4850 & 0.0000 \\ 
      147.1201 & 0.5199 & 0.0000 \\ 
      157.9072 & 0.5494 & 0.0000 \\ 
      169.2090 & 0.5759 & 0.0000 \\ 
      181.0377 & 0.5993 & 0.0000 \\ 
      193.4052 & 0.6207 & 0.0000 \\ 
      206.3234 & 0.6403 & 0.0000 \\ 
      219.8045 & 0.6585 & 0.0000 \\ 
      233.8603 & 0.6751 & 0.0000 \\ 
      248.5030 & 0.6908 & 0.0000 \\ 
\end{longtable} 
 
\clearpage

\begin{longtable}{d{4}d{4}d{4}} 
  \caption{ Raw data for Cu as calculated with LSDA in VASP. \label{tab:cu_main_lsd_v_raw_dat} } \\ \toprule 
       \multicolumn{1}{l}{$V_{\text{prim}}$ (\AA{}$^3$)} & \multicolumn{1}{l}{CFR fraction} & \multicolumn{1}{l}{Band gap from DOS (eV)} \\ \hline 
\endfirsthead 
\toprule
       \multicolumn{1}{l}{$V_{\text{prim}}$ (\AA{}$^3$)} & \multicolumn{1}{l}{CFR fraction} & \multicolumn{1}{l}{Band gap from DOS (eV)} \\ \hline 
\multicolumn{1}{c}{$\vdots$}& \multicolumn{1}{c}{$\vdots$}& \multicolumn{1}{c}{$\vdots$}\\ 
\endhead 
\multicolumn{1}{c}{$\vdots$}& \multicolumn{1}{c}{$\vdots$}& \multicolumn{1}{c}{$\vdots$}\\ 
\hline \hline 
\endfoot 
\hline \hline 
\endlastfoot
      10.9036 & 0.0000 & 0.0000 \\ 
      12.8697 & 0.0000 & 0.0000 \\ 
      15.0591 & 0.0000 & 0.0000 \\ 
      17.4836 & 0.0000 & 0.0000 \\ 
      20.1554 & 0.0000 & 0.0000 \\ 
      23.0864 & 0.0000 & 0.0000 \\ 
      26.2885 & 0.0000 & 0.0000 \\ 
      29.7739 & 0.0020 & 0.0000 \\ 
      33.5544 & 0.0225 & 0.0000 \\ 
      37.6422 & 0.0562 & 0.0000 \\ 
      42.0492 & 0.1090 & 0.0000 \\ 
      46.7873 & 0.2345 & 0.0000 \\ 
      51.8687 & 0.3557 & 0.0000 \\ 
      57.3052 & 0.4634 & 0.0000 \\ 
      63.1090 & 0.5328 & 0.0000 \\ 
      69.2920 & 0.5834 & 0.0000 \\ 
      75.8661 & 0.6249 & 0.0000 \\ 
      82.8435 & 0.6593 & 0.0000 \\ 
      90.2360 & 0.6889 & 0.0000 \\ 
      98.0558 & 0.7143 & 0.0000 \\ 
      106.3148 & 0.7367 & 0.0000 \\ 
      115.0249 & 0.7564 & 0.0000 \\ 
      124.1983 & 0.7742 & 0.0000 \\ 
      133.8468 & 0.7902 & 0.0000 \\ 
      143.9826 & 0.8047 & 0.0000 \\ 
      154.6176 & 0.8180 & 0.0000 \\ 
      165.7637 & 0.8299 & 0.0000 \\ 
      177.4331 & 0.8408 & 0.0000 \\ 
      189.6376 & 0.8506 & 0.0000 \\ 
      202.3894 & 0.8600 & 0.0000 \\ 
      215.7004 & 0.8685 & 0.0000 \\ 
      229.5825 & 0.8763 & 0.0000 \\ 
      244.0479 & 0.8835 & 0.0000 \\ 
\end{longtable} 
 
\clearpage

\begin{longtable}{d{4}d{4}d{4}} 
  \caption{ Raw data for Pt as calculated with LSDA in VASP. \label{tab:pt_main_lsd_v_raw_dat} } \\ \toprule 
       \multicolumn{1}{l}{$V_{\text{prim}}$ (\AA{}$^3$)} & \multicolumn{1}{l}{CFR fraction} & \multicolumn{1}{l}{Band gap from DOS (eV)} \\ \hline 
\endfirsthead 
\toprule
       \multicolumn{1}{l}{$V_{\text{prim}}$ (\AA{}$^3$)} & \multicolumn{1}{l}{CFR fraction} & \multicolumn{1}{l}{Band gap from DOS (eV)} \\ \hline 
\multicolumn{1}{c}{$\vdots$}& \multicolumn{1}{c}{$\vdots$}& \multicolumn{1}{c}{$\vdots$}\\ 
\endhead 
\multicolumn{1}{c}{$\vdots$}& \multicolumn{1}{c}{$\vdots$}& \multicolumn{1}{c}{$\vdots$}\\ 
\hline \hline 
\endfoot 
\hline \hline 
\endlastfoot
      14.8297 & 0.0000 & 0.0000 \\ 
      17.2302 & 0.0000 & 0.0000 \\ 
      19.8767 & 0.0000 & 0.0000 \\ 
      22.7813 & 0.0000 & 0.0000 \\ 
      25.9558 & 0.0052 & 0.0000 \\ 
      29.4123 & 0.0314 & 0.0000 \\ 
      33.1627 & 0.0724 & 0.0000 \\ 
      37.2193 & 0.1653 & 0.0000 \\ 
      41.5938 & 0.2868 & 0.0000 \\ 
      46.2983 & 0.3966 & 0.0000 \\ 
      51.3447 & 0.4821 & 0.0000 \\ 
      56.7452 & 0.5396 & 0.0000 \\ 
      62.5117 & 0.5843 & 0.0000 \\ 
      68.6562 & 0.6213 & 0.0000 \\ 
      75.1907 & 0.6525 & 0.0000 \\ 
      82.1273 & 0.6802 & 0.0000 \\ 
      89.4778 & 0.7042 & 0.0000 \\ 
      97.2542 & 0.7253 & 0.0000 \\ 
      105.4688 & 0.7445 & 0.0000 \\ 
      114.1333 & 0.7620 & 0.0000 \\ 
      123.2598 & 0.7777 & 0.0000 \\ 
      132.8603 & 0.7928 & 0.0000 \\ 
      142.9468 & 0.8059 & 0.0000 \\ 
      153.5312 & 0.8182 & 0.0000 \\ 
      164.6258 & 0.8295 & 0.0000 \\ 
      176.2422 & 0.8400 & 0.0000 \\ 
      188.3928 & 0.8494 & 0.0000 \\ 
      201.0893 & 0.8584 & 0.0000 \\ 
      214.3438 & 0.8667 & 0.0000 \\ 
      228.1683 & 0.8746 & 0.0000 \\ 
      242.5748 & 0.8817 & 0.0000 \\ 
\end{longtable} 
 
\clearpage

\begin{longtable}{d{4}d{4}d{4}} 
  \caption{ Raw data for C as calculated with LSDA in VASP. \label{tab:c_main_lsd_v_raw_dat} } \\ \toprule 
       \multicolumn{1}{l}{$V_{\text{prim}}$ (\AA{}$^3$)} & \multicolumn{1}{l}{CFR fraction} & \multicolumn{1}{l}{Band gap from DOS (eV)} \\ \hline 
\endfirsthead 
\toprule
       \multicolumn{1}{l}{$V_{\text{prim}}$ (\AA{}$^3$)} & \multicolumn{1}{l}{CFR fraction} & \multicolumn{1}{l}{Band gap from DOS (eV)} \\ \hline 
\multicolumn{1}{c}{$\vdots$}& \multicolumn{1}{c}{$\vdots$}& \multicolumn{1}{c}{$\vdots$}\\ 
\endhead 
\multicolumn{1}{c}{$\vdots$}& \multicolumn{1}{c}{$\vdots$}& \multicolumn{1}{c}{$\vdots$}\\ 
\hline \hline 
\endfoot 
\hline \hline 
\endlastfoot
      10.9967 & 0.0000 & 4.0588 \\ 
      12.9738 & 0.0000 & 3.7563 \\ 
      15.1746 & 0.0000 & 3.3818 \\ 
      17.6112 & 0.0000 & 3.0743 \\ 
      20.2957 & 0.0000 & 2.5228 \\ 
      23.2399 & 0.0097 & 1.0881 \\ 
      26.4560 & 0.0827 & 0.0000 \\ 
      29.9558 & 0.1643 & 0.0000 \\ 
      33.7514 & 0.2401 & 0.0000 \\ 
      37.8549 & 0.3093 & 0.0000 \\ 
      42.2781 & 0.3676 & 0.0000 \\ 
      47.0331 & 0.4171 & 0.0000 \\ 
      52.1320 & 0.4621 & 0.0000 \\ 
      57.5866 & 0.5012 & 0.0000 \\ 
      63.4090 & 0.5374 & 0.0000 \\ 
      69.6113 & 0.5703 & 0.0000 \\ 
      76.2053 & 0.6005 & 0.0000 \\ 
      83.2031 & 0.6294 & 0.0000 \\ 
      90.6168 & 0.6555 & 0.0000 \\ 
      98.4582 & 0.6803 & 0.0000 \\ 
      106.7394 & 0.7041 & 0.0000 \\ 
      115.4725 & 0.7258 & 0.0000 \\ 
      124.6693 & 0.7471 & 0.0000 \\ 
      134.3419 & 0.7665 & 0.0000 \\ 
      144.5024 & 0.7837 & 0.0000 \\ 
      155.1626 & 0.7981 & 0.0000 \\ 
      166.3347 & 0.8106 & 0.0000 \\ 
      178.0305 & 0.8224 & 0.0000 \\ 
      190.2621 & 0.8327 & 0.0000 \\ 
      203.0416 & 0.8422 & 0.0000 \\ 
      216.3808 & 0.8512 & 0.0000 \\ 
      230.2918 & 0.8591 & 0.0000 \\ 
      244.7867 & 0.8666 & 0.0000 \\ 
\end{longtable} 
 
\clearpage

\begin{longtable}{d{4}d{4}d{4}} 
  \caption{ Raw data for Ne as calculated with LSDA in VASP. \label{tab:ne_main_lsd_v_raw_dat} } \\ \toprule 
       \multicolumn{1}{l}{$V_{\text{prim}}$ (\AA{}$^3$)} & \multicolumn{1}{l}{CFR fraction} & \multicolumn{1}{l}{Band gap from DOS (eV)} \\ \hline 
\endfirsthead 
\toprule
       \multicolumn{1}{l}{$V_{\text{prim}}$ (\AA{}$^3$)} & \multicolumn{1}{l}{CFR fraction} & \multicolumn{1}{l}{Band gap from DOS (eV)} \\ \hline 
\multicolumn{1}{c}{$\vdots$}& \multicolumn{1}{c}{$\vdots$}& \multicolumn{1}{c}{$\vdots$}\\ 
\endhead 
\multicolumn{1}{c}{$\vdots$}& \multicolumn{1}{c}{$\vdots$}& \multicolumn{1}{c}{$\vdots$}\\ 
\hline \hline 
\endfoot 
\hline \hline 
\endlastfoot
      5.7873 & 0.0261 & 18.3436 \\ 
      6.0972 & 0.0576 & 17.6710 \\ 
      6.4181 & 0.1407 & 17.0007 \\ 
      6.7500 & 0.2294 & 16.4522 \\ 
      8.1920 & 0.5164 & 14.4664 \\ 
      9.8260 & 0.6455 & 13.4261 \\ 
      11.6640 & 0.7178 & 12.4733 \\ 
      13.7180 & 0.7691 & 11.9869 \\ 
      16.0000 & 0.8060 & 11.6004 \\ 
      18.5220 & 0.8354 & 11.3754 \\ 
      21.2960 & 0.8570 & 11.3931 \\ 
      24.3340 & 0.8745 & 11.3476 \\ 
      27.6480 & 0.8899 & 11.2333 \\ 
      31.2500 & 0.9024 & 11.3534 \\ 
      35.1520 & 0.9131 & 11.3287 \\ 
      39.3660 & 0.9227 & 11.4308 \\ 
      43.9040 & 0.9306 & 11.5933 \\ 
      48.7780 & 0.9373 & 11.6459 \\ 
      54.0000 & 0.9436 & 11.7969 \\ 
\end{longtable} 
 
\clearpage

\begin{longtable}{d{4}d{4}d{4}} 
  \caption{ Raw data for NaCl as calculated with LSDA in VASP. \label{tab:nacl_main_lsd_v_raw_dat} } \\ \toprule 
       \multicolumn{1}{l}{$V_{\text{prim}}$ (\AA{}$^3$)} & \multicolumn{1}{l}{CFR fraction} & \multicolumn{1}{l}{Band gap from DOS (eV)} \\ \hline 
\endfirsthead 
\toprule
       \multicolumn{1}{l}{$V_{\text{prim}}$ (\AA{}$^3$)} & \multicolumn{1}{l}{CFR fraction} & \multicolumn{1}{l}{Band gap from DOS (eV)} \\ \hline 
\multicolumn{1}{c}{$\vdots$}& \multicolumn{1}{c}{$\vdots$}& \multicolumn{1}{c}{$\vdots$}\\ 
\endhead 
\multicolumn{1}{c}{$\vdots$}& \multicolumn{1}{c}{$\vdots$}& \multicolumn{1}{c}{$\vdots$}\\ 
\hline \hline 
\endfoot 
\hline \hline 
\endlastfoot
      10.4455 & 0.0000 & 0.0000 \\ 
      12.3577 & 0.0000 & 0.0000 \\ 
      14.4902 & 0.0000 & 1.7634 \\ 
      16.8548 & 0.0000 & 3.2939 \\ 
      19.4636 & 0.0000 & 4.4587 \\ 
      22.3287 & 0.0000 & 5.4665 \\ 
      25.4619 & 0.0000 & 6.1151 \\ 
      28.8753 & 0.0000 & 6.7088 \\ 
      32.5810 & 0.0286 & 6.2297 \\ 
      36.5908 & 0.0884 & 5.5120 \\ 
      40.9168 & 0.1780 & 5.0743 \\ 
      45.5711 & 0.2580 & 4.5858 \\ 
      50.5655 & 0.3391 & 4.2189 \\ 
      55.9121 & 0.4023 & 3.9347 \\ 
      61.6230 & 0.4586 & 3.7337 \\ 
      67.7100 & 0.5119 & 3.5726 \\ 
      74.1852 & 0.5587 & 3.4197 \\ 
      81.0607 & 0.5957 & 3.1547 \\ 
      88.3483 & 0.6348 & 3.0390 \\ 
      96.0601 & 0.6682 & 2.9438 \\ 
      104.2082 & 0.6958 & 2.8269 \\ 
      112.8044 & 0.7157 & 2.7118 \\ 
      121.8609 & 0.7331 & 2.5950 \\ 
      131.3895 & 0.7504 & 2.4746 \\ 
      141.4023 & 0.7623 & 2.3570 \\ 
      151.9114 & 0.7769 & 2.3220 \\ 
      162.9286 & 0.7869 & 2.2251 \\ 
      174.4660 & 0.7982 & 2.1907 \\ 
      186.5357 & 0.8045 & 2.0956 \\ 
      199.1495 & 0.8142 & 1.9712 \\ 
      212.3195 & 0.8224 & 1.9687 \\ 
      226.0578 & 0.8293 & 1.8462 \\ 
\end{longtable} 
 
\clearpage

\begin{longtable}{d{4}d{4}d{4}} 
  \caption{ Raw data for Si as calculated with LSDA in VASP. \label{tab:si_main_lsd_v_raw_dat} } \\ \toprule 
       \multicolumn{1}{l}{$V_{\text{prim}}$ (\AA{}$^3$)} & \multicolumn{1}{l}{CFR fraction} & \multicolumn{1}{l}{Band gap from DOS (eV)} \\ \hline 
\endfirsthead 
\toprule
       \multicolumn{1}{l}{$V_{\text{prim}}$ (\AA{}$^3$)} & \multicolumn{1}{l}{CFR fraction} & \multicolumn{1}{l}{Band gap from DOS (eV)} \\ \hline 
\multicolumn{1}{c}{$\vdots$}& \multicolumn{1}{c}{$\vdots$}& \multicolumn{1}{c}{$\vdots$}\\ 
\endhead 
\multicolumn{1}{c}{$\vdots$}& \multicolumn{1}{c}{$\vdots$}& \multicolumn{1}{c}{$\vdots$}\\ 
\hline \hline 
\endfoot 
\hline \hline 
\endlastfoot
      39.3660 & 0.0000 & 0.3872 \\ 
      43.9040 & 0.0000 & 0.6069 \\ 
      48.7780 & 0.0000 & 0.6498 \\ 
      54.0000 & 0.0105 & 0.1719 \\ 
      59.5820 & 0.0497 & 0.0000 \\ 
      65.5360 & 0.1094 & 0.0000 \\ 
      71.8740 & 0.1685 & 0.0000 \\ 
      78.6080 & 0.2202 & 0.0000 \\ 
      85.7500 & 0.2672 & 0.0000 \\ 
      93.3120 & 0.3090 & 0.0000 \\ 
      101.3060 & 0.3468 & 0.0000 \\ 
      109.7440 & 0.3819 & 0.0000 \\ 
      118.6380 & 0.4135 & 0.0000 \\ 
      128.0000 & 0.4430 & 0.0000 \\ 
      137.8420 & 0.4716 & 0.0000 \\ 
      148.1760 & 0.4970 & 0.0000 \\ 
      159.0140 & 0.5221 & 0.0000 \\ 
      170.3680 & 0.5455 & 0.0000 \\ 
      182.2500 & 0.5678 & 0.0000 \\ 
      194.6720 & 0.5892 & 0.0000 \\ 
      207.6460 & 0.6091 & 0.0000 \\ 
      221.1840 & 0.6285 & 0.0000 \\ 
      235.2980 & 0.6472 & 0.0000 \\ 
      250.0000 & 0.6648 & 0.0000 \\ 
\end{longtable} 
 
\clearpage

\begin{longtable}{d{4}d{4}} 
  \caption{ Raw data for Al as calculated with PBE in Castep. \label{tab:al_main_pbe_c_raw_dat} } \\ \toprule 
       \multicolumn{1}{l}{$V_{\text{prim}}$ (\AA{}$^3$)} & \multicolumn{1}{l}{CFR fraction} \\ \hline 
\endfirsthead 
\toprule
       \multicolumn{1}{l}{$V_{\text{prim}}$ (\AA{}$^3$)} & \multicolumn{1}{l}{CFR fraction} \\ \hline 
\multicolumn{1}{c}{$\vdots$}& \multicolumn{1}{c}{$\vdots$}\\ 
\endhead 
\multicolumn{1}{c}{$\vdots$}& \multicolumn{1}{c}{$\vdots$}\\ 
\hline \hline 
\endfoot 
\hline \hline 
\endlastfoot
      3.9062 & 0.0000 \\ 
      4.3940 & 0.0000 \\ 
      4.9207 & 0.0000 \\ 
      5.4880 & 0.0000 \\ 
      6.0972 & 0.0000 \\ 
      6.7500 & 0.0000 \\ 
      7.4478 & 0.0000 \\ 
      8.1920 & 0.0000 \\
\end{longtable} 
 
\clearpage

\begin{longtable}{d{4}d{4}} 
  \caption{ Raw data for Cu as calculated with PBE in Castep. \label{tab:cu_main_pbe_c_raw_dat} } \\ \toprule 
       \multicolumn{1}{l}{$V_{\text{prim}}$ (\AA{}$^3$)} & \multicolumn{1}{l}{CFR fraction} \\ \hline 
\endfirsthead 
\toprule
       \multicolumn{1}{l}{$V_{\text{prim}}$ (\AA{}$^3$)} & \multicolumn{1}{l}{CFR fraction} \\ \hline 
\multicolumn{1}{c}{$\vdots$}& \multicolumn{1}{c}{$\vdots$}\\ 
\endhead 
\multicolumn{1}{c}{$\vdots$}& \multicolumn{1}{c}{$\vdots$}\\ 
\hline \hline 
\endfoot 
\hline \hline 
\endlastfoot
      3.9062 & 0.0000 \\ 
      4.3940 & 0.0000 \\ 
      4.9207 & 0.0000 \\ 
      5.4880 & 0.0000 \\ 
      6.0972 & 0.0000 \\ 
      6.7500 & 0.0000 \\ 
      7.4478 & 0.0000 \\ 
      8.1920 & 0.0000 \\ 
      8.9842 & 0.0000 \\ 
      9.8260 & 0.0000 \\ 
      10.7188 & 0.0000 \\ 
      11.6640 & 0.0000 \\ 
      12.6633 & 0.0000 \\ 
      13.7180 & 0.0000 \\ 
      14.8297 & 0.0000 \\ 
      16.0000 & 0.0000 \\ 
      17.2302 & 0.0000 \\ 
      18.5220 & 0.0000 \\ 
      19.8767 & 0.0000 \\ 
      21.2960 & 0.0000 \\ 
      22.7812 & 0.0000 \\ 
      24.3340 & 0.0000 \\ 
      25.9558 & 0.0000 \\ 
      27.6480 & 0.0000 \\ 
      29.4123 & 0.0000 \\ 
      31.2500 & 0.0000 \\ 
      33.1627 & 0.0019 \\ 
      35.1520 & 0.0095 \\ 
      37.2192 & 0.0239 \\ 
      39.3660 & 0.0448 \\ 
      41.5938 & 0.0718 \\ 
      43.9040 & 0.0959 \\ 
      46.2983 & 0.1387 \\ 
      48.7780 & 0.2028 \\ 
      51.3448 & 0.3226 \\ 
      54.0000 & 0.4044 \\ 
      56.7452 & 0.4711 \\ 
      59.5820 & 0.5177 \\ 
      62.5117 & 0.5479 \\ 
      65.5360 & 0.5663 \\ 
      68.6562 & 0.5832 \\ 
      71.8740 & 0.6008 \\ 
      75.1908 & 0.6181 \\ 
      78.6080 & 0.6329 \\ 
      82.1273 & 0.6500 \\ 
\end{longtable} 
 
\clearpage

\begin{longtable}{d{4}d{4}} 
  \caption{ Raw data for Pt as calculated with PBE in Castep. \label{tab:pt_main_pbe_c_raw_dat} } \\ \toprule 
       \multicolumn{1}{l}{$V_{\text{prim}}$ (\AA{}$^3$)} & \multicolumn{1}{l}{CFR fraction} \\ \hline 
\endfirsthead 
\toprule
       \multicolumn{1}{l}{$V_{\text{prim}}$ (\AA{}$^3$)} & \multicolumn{1}{l}{CFR fraction} \\ \hline 
\multicolumn{1}{c}{$\vdots$}& \multicolumn{1}{c}{$\vdots$}\\ 
\endhead 
\multicolumn{1}{c}{$\vdots$}& \multicolumn{1}{c}{$\vdots$}\\ 
\hline \hline 
\endfoot 
\hline \hline 
\endlastfoot
      4.3940 & 0.0000 \\ 
      4.9207 & 0.0000 \\ 
      5.4880 & 0.0000 \\ 
      6.0972 & 0.0000 \\ 
      6.7500 & 0.0000 \\ 
      7.4478 & 0.0000 \\ 
      8.1920 & 0.0000 \\ 
      8.9842 & 0.0000 \\ 
      9.8260 & 0.0000 \\ 
      10.7188 & 0.0000 \\ 
      11.6640 & 0.0000 \\ 
      12.6633 & 0.0000 \\ 
      13.7180 & 0.0000 \\ 
      14.8297 & 0.0000 \\ 
      16.0000 & 0.0000 \\ 
      17.2302 & 0.0000 \\ 
      18.5220 & 0.0000 \\ 
      19.8767 & 0.0000 \\ 
      21.2960 & 0.0000 \\ 
      22.7812 & 0.0000 \\ 
      24.3340 & 0.0000 \\ 
      25.9558 & 0.0000 \\ 
      27.6480 & 0.0040 \\ 
      29.4123 & 0.0209 \\ 
      31.2500 & 0.0402 \\ 
      33.1627 & 0.0680 \\ 
      35.1520 & 0.0870 \\ 
      37.2192 & 0.1105 \\ 
      39.3660 & 0.2034 \\ 
      41.5938 & 0.2942 \\ 
      43.9040 & 0.3549 \\ 
      46.2983 & 0.4063 \\ 
      48.7780 & 0.4611 \\ 
      51.3448 & 0.4893 \\ 
      54.0000 & 0.5179 \\ 
      56.7452 & 0.5427 \\ 
      59.5820 & 0.5593 \\ 
      62.5117 & 0.5743 \\ 
      65.5360 & 0.5953 \\ 
      68.6562 & 0.6127 \\ 
      71.8740 & 0.6310 \\ 
      75.1908 & 0.6457 \\ 
      78.6080 & 0.6598 \\ 
      82.1273 & 0.6692 \\ 
      85.7500 & 0.6799 \\ 
      89.4777 & 0.6891 \\ 
\end{longtable} 
 
\clearpage

\twocolumngrid
\begin{longtable}{d{4}d{4}} 
  \caption{ Raw data for C as calculated with PBE in Castep. \label{tab:c_main_pbe_c_raw_dat} } \\ \toprule 
       \multicolumn{1}{l}{$V_{\text{prim}}$ (\AA{}$^3$)} & \multicolumn{1}{l}{CFR fraction} \\ \hline 
\endfirsthead 
\toprule
       \multicolumn{1}{l}{$V_{\text{prim}}$ (\AA{}$^3$)} & \multicolumn{1}{l}{CFR fraction} \\ \hline 
\multicolumn{1}{c}{$\vdots$}& \multicolumn{1}{c}{$\vdots$}\\ 
\endhead 
\multicolumn{1}{c}{$\vdots$}& \multicolumn{1}{c}{$\vdots$}\\ 
\hline \hline 
\endfoot 
\hline \hline 
\endlastfoot
      3.9062 & 0.0000 \\ 
      4.3940 & 0.0000 \\ 
      4.9208 & 0.0000 \\ 
      5.4880 & 0.0000 \\ 
      6.0972 & 0.0000 \\ 
      6.7500 & 0.0000 \\ 
      7.4478 & 0.0000 \\ 
      8.1920 & 0.0000 \\
\end{longtable} 
 
\clearpage

\begin{longtable}{d{4}d{4}} 
  \caption{ Raw data for Ne as calculated with PBE in Castep. \label{tab:ne_main_pbe_c_raw_dat} } \\ \toprule 
       \multicolumn{1}{l}{$V_{\text{prim}}$ (\AA{}$^3$)} & \multicolumn{1}{l}{CFR fraction} \\ \hline 
\endfirsthead 
\toprule
       \multicolumn{1}{l}{$V_{\text{prim}}$ (\AA{}$^3$)} & \multicolumn{1}{l}{CFR fraction} \\ \hline 
\multicolumn{1}{c}{$\vdots$}& \multicolumn{1}{c}{$\vdots$}\\ 
\endhead 
\multicolumn{1}{c}{$\vdots$}& \multicolumn{1}{c}{$\vdots$}\\ 
\hline \hline 
\endfoot 
\hline \hline 
\endlastfoot
      3.9062 & 0.0000 \\ 
      4.3940 & 0.0000 \\ 
      4.9207 & 0.0000 \\ 
      5.4880 & 0.0359 \\ 
      6.0972 & 0.1041 \\ 
      6.7500 & 0.2904 \\ 
      7.4478 & 0.4156 \\ 
      8.1920 & 0.4829 \\
\end{longtable} 
 
\clearpage

\begin{longtable}{d{4}d{4}} 
  \caption{ Raw data for NaCl as calculated with PBE in Castep. \label{tab:nacl_main_pbe_c_raw_dat} } \\ \toprule 
       \multicolumn{1}{l}{$V_{\text{prim}}$ (\AA{}$^3$)} & \multicolumn{1}{l}{CFR fraction} \\ \hline 
\endfirsthead 
\toprule
       \multicolumn{1}{l}{$V_{\text{prim}}$ (\AA{}$^3$)} & \multicolumn{1}{l}{CFR fraction} \\ \hline 
\multicolumn{1}{c}{$\vdots$}& \multicolumn{1}{c}{$\vdots$}\\ 
\endhead 
\multicolumn{1}{c}{$\vdots$}& \multicolumn{1}{c}{$\vdots$}\\ 
\hline \hline 
\endfoot 
\hline \hline 
\endlastfoot
      4.3940 & 0.0000 \\ 
      4.9208 & 0.0000 \\ 
      5.4880 & 0.0000 \\ 
      6.0972 & 0.0000 \\ 
      6.7500 & 0.0000 \\ 
      7.4478 & 0.0000 \\ 
      8.1920 & 0.0000 \\ 
      8.9842 & 0.0000 \\
\end{longtable} 
 
\clearpage

\begin{longtable}{d{4}d{4}} 
  \caption{ Raw data for Si as calculated with PBE in Castep. \label{tab:si_main_pbe_c_raw_dat} } \\ \toprule 
       \multicolumn{1}{l}{$V_{\text{prim}}$ (\AA{}$^3$)} & \multicolumn{1}{l}{CFR fraction} \\ \hline 
\endfirsthead 
\toprule
       \multicolumn{1}{l}{$V_{\text{prim}}$ (\AA{}$^3$)} & \multicolumn{1}{l}{CFR fraction} \\ \hline 
\multicolumn{1}{c}{$\vdots$}& \multicolumn{1}{c}{$\vdots$}\\ 
\endhead 
\multicolumn{1}{c}{$\vdots$}& \multicolumn{1}{c}{$\vdots$}\\ 
\hline \hline 
\endfoot 
\hline \hline 
\endlastfoot
      3.9062 & 0.0000 \\ 
      4.3940 & 0.0000 \\ 
      4.9208 & 0.0000 \\ 
      5.4880 & 0.0000 \\ 
      6.0972 & 0.0000 \\ 
      6.7500 & 0.0000 \\ 
      7.4478 & 0.0000 \\ 
      8.1920 & 0.0000 \\
\end{longtable} 
 
\clearpage

\clearpage

\onecolumngrid
\section{Raw data for Group 1 elemental solids}

\begin{longtable}{d{4}d{4}} 
  \caption{ Raw data for Li as calculated with PBE in Castep. \label{tab:li_group_1_pbe_c_raw_dat} } \\ \toprule 
       \multicolumn{1}{l}{$V_{\text{prim}}$ (\AA{}$^3$)} & \multicolumn{1}{l}{CFR fraction} \\ \hline 
\endfirsthead 
\toprule
       \multicolumn{1}{l}{$V_{\text{prim}}$ (\AA{}$^3$)} & \multicolumn{1}{l}{CFR fraction} \\ \hline 
\multicolumn{1}{c}{$\vdots$}& \multicolumn{1}{c}{$\vdots$}\\ 
\endhead 
\multicolumn{1}{c}{$\vdots$}& \multicolumn{1}{c}{$\vdots$}\\ 
\hline \hline 
\endfoot 
\hline \hline 
\endlastfoot
      4.0000 & 0.0000 \\ 
      4.6305 & 0.0000 \\ 
      5.3240 & 0.0000 \\ 
      6.0835 & 0.0000 \\ 
      6.9120 & 0.0000 \\ 
      7.8125 & 0.0000 \\ 
      8.7880 & 0.0000 \\ 
      9.8415 & 0.0000 \\
\end{longtable} 
 
\clearpage

\begin{longtable}{d{4}d{4}} 
  \caption{ Raw data for Na as calculated with PBE in Castep. \label{tab:na_group_1_pbe_c_raw_dat} } \\ \toprule 
       \multicolumn{1}{l}{$V_{\text{prim}}$ (\AA{}$^3$)} & \multicolumn{1}{l}{CFR fraction} \\ \hline 
\endfirsthead 
\toprule
       \multicolumn{1}{l}{$V_{\text{prim}}$ (\AA{}$^3$)} & \multicolumn{1}{l}{CFR fraction} \\ \hline 
\multicolumn{1}{c}{$\vdots$}& \multicolumn{1}{c}{$\vdots$}\\ 
\endhead 
\multicolumn{1}{c}{$\vdots$}& \multicolumn{1}{c}{$\vdots$}\\ 
\hline \hline 
\endfoot 
\hline \hline 
\endlastfoot
      4.0000 & 0.0000 \\ 
      4.6305 & 0.0000 \\ 
      5.3240 & 0.0000 \\ 
      6.0835 & 0.0000 \\ 
      6.9120 & 0.0000 \\ 
      7.8125 & 0.0000 \\ 
      8.7880 & 0.0000 \\ 
      9.8415 & 0.0000 \\
\end{longtable} 
 
\clearpage

\begin{longtable}{d{4}d{4}} 
  \caption{ Raw data for K as calculated with PBE in Castep. \label{tab:k_group_1_pbe_c_raw_dat} } \\ \toprule 
       \multicolumn{1}{l}{$V_{\text{prim}}$ (\AA{}$^3$)} & \multicolumn{1}{l}{CFR fraction} \\ \hline 
\endfirsthead 
\toprule
       \multicolumn{1}{l}{$V_{\text{prim}}$ (\AA{}$^3$)} & \multicolumn{1}{l}{CFR fraction} \\ \hline 
\multicolumn{1}{c}{$\vdots$}& \multicolumn{1}{c}{$\vdots$}\\ 
\endhead 
\multicolumn{1}{c}{$\vdots$}& \multicolumn{1}{c}{$\vdots$}\\ 
\hline \hline 
\endfoot 
\hline \hline 
\endlastfoot
      105.2869 & 0.0000 \\ 
      111.6025 & 0.0000 \\ 
      118.1657 & 0.0000 \\ 
      124.9814 & 0.0000 \\ 
      132.0541 & 0.0000 \\ 
      139.3888 & 0.0000 \\ 
      146.9902 & 0.0000 \\ 
      154.8631 & 0.0000 \\
\end{longtable} 
 
\clearpage

\begin{longtable}{d{4}d{4}} 
  \caption{ Raw data for Rb as calculated with PBE in Castep. \label{tab:rb_group_1_pbe_c_raw_dat} } \\ \toprule 
       \multicolumn{1}{l}{$V_{\text{prim}}$ (\AA{}$^3$)} & \multicolumn{1}{l}{CFR fraction} \\ \hline 
\endfirsthead 
\toprule
       \multicolumn{1}{l}{$V_{\text{prim}}$ (\AA{}$^3$)} & \multicolumn{1}{l}{CFR fraction} \\ \hline 
\multicolumn{1}{c}{$\vdots$}& \multicolumn{1}{c}{$\vdots$}\\ 
\endhead 
\multicolumn{1}{c}{$\vdots$}& \multicolumn{1}{c}{$\vdots$}\\ 
\hline \hline 
\endfoot 
\hline \hline 
\endlastfoot
      105.2869 & 0.0000 \\ 
      111.6025 & 0.0000 \\ 
      118.1657 & 0.0000 \\ 
      124.9814 & 0.0000 \\ 
      132.0541 & 0.0000 \\ 
      139.3888 & 0.0000 \\ 
      146.9902 & 0.0000 \\ 
      154.8631 & 0.0000 \\
\end{longtable} 
 
\clearpage

\onecolumngrid
\begin{longtable}{d{4}d{4}} 
  \caption{ Raw data for Cs as calculated with PBE in Castep. \label{tab:cs_group_1_pbe_c_raw_dat} } \\ \toprule 
       \multicolumn{1}{l}{$V_{\text{prim}}$ (\AA{}$^3$)} & \multicolumn{1}{l}{CFR fraction} \\ \hline 
\endfirsthead 
\toprule
       \multicolumn{1}{l}{$V_{\text{prim}}$ (\AA{}$^3$)} & \multicolumn{1}{l}{CFR fraction} \\ \hline 
\multicolumn{1}{c}{$\vdots$}& \multicolumn{1}{c}{$\vdots$}\\ 
\endhead 
\multicolumn{1}{c}{$\vdots$}& \multicolumn{1}{c}{$\vdots$}\\ 
\hline \hline 
\endfoot 
\hline \hline 
\endlastfoot
      105.2869 & 0.0000 \\ 
      111.6025 & 0.0000 \\ 
      118.1657 & 0.0000 \\ 
      124.9814 & 0.0000 \\ 
      132.0541 & 0.0000 \\ 
      139.3888 & 0.0000 \\ 
      146.9902 & 0.0000 \\ 
      154.8631 & 0.0000 \\ 
      163.0121 & 0.0000 \\ 
      171.4422 & 0.0000 \\ 
      180.1579 & 0.0000 \\ 
      189.1642 & 0.0000 \\ 
      198.4658 & 0.0000 \\ 
      208.0673 & 0.0000 \\ 
      217.9736 & 0.0000 \\ 
      228.1895 & 0.0000 \\ 
      238.7197 & 0.0000 \\ 
      249.5690 & 0.0000 \\ 
      260.7421 & 0.0000 \\ 
      272.2438 & 0.0000 \\ 
      284.0789 & 0.0000 \\ 
      296.2521 & 0.0000 \\ 
      308.7682 & 0.0000 \\ 
      321.6319 & 0.0000 \\ 
      334.8480 & 0.0008 \\ 
      348.4213 & 0.0133 \\ 
      362.3565 & 0.0649 \\ 
      376.6585 & 0.1087 \\ 
      391.3318 & 0.1452 \\ 
      421.8120 & 0.2068 \\ 
      437.6284 & 0.2343 \\ 
      453.8352 & 0.2599 \\ 
      470.4373 & 0.2852 \\ 
      487.4395 & 0.3100 \\ 
      504.8464 & 0.3316 \\ 
      522.6629 & 0.3536 \\ 
      540.8937 & 0.3743 \\ 
      559.5436 & 0.3944 \\ 
      578.6173 & 0.4154 \\ 
      598.1197 & 0.4353 \\ 
      618.0554 & 0.4558 \\ 
      638.4292 & 0.4744 \\ 
      659.2460 & 0.4849 \\ 
      680.5104 & 0.5138 \\ 
      702.2272 & 0.5325 \\ 
      747.0371 & 0.5648 \\ 
      770.1397 & 0.5818 \\ 
      793.7138 & 0.5968 \\ 
      817.7633 & 0.6089 \\ 
      842.2970 & 0.6208 \\ 
\end{longtable} 
 
\clearpage

\clearpage

\onecolumngrid
\section{Raw data for Group 2 elemental solids}

\twocolumngrid
\begin{longtable}{d{4}d{4}} 
  \caption{ Raw data for Be as calculated with PBE in Castep. \label{tab:be_group_2_pbe_c_raw_dat} } \\ \toprule 
       \multicolumn{1}{l}{$V_{\text{prim}}$ (\AA{}$^3$)} & \multicolumn{1}{l}{CFR fraction} \\ \hline 
\endfirsthead 
\toprule
       \multicolumn{1}{l}{$V_{\text{prim}}$ (\AA{}$^3$)} & \multicolumn{1}{l}{CFR fraction} \\ \hline 
\multicolumn{1}{c}{$\vdots$}& \multicolumn{1}{c}{$\vdots$}\\ 
\endhead 
\multicolumn{1}{c}{$\vdots$}& \multicolumn{1}{c}{$\vdots$}\\ 
\hline \hline 
\endfoot 
\hline \hline 
\endlastfoot
      4.0000 & 0.0000 \\ 
      4.6305 & 0.0000 \\ 
      5.3240 & 0.0000 \\ 
      6.0835 & 0.0000 \\ 
      6.9120 & 0.0000 \\ 
      7.8125 & 0.0000 \\ 
      8.7880 & 0.0000 \\ 
      9.8415 & 0.0000 \\
\end{longtable} 
 
\clearpage

\begin{longtable}{d{4}d{4}} 
  \caption{ Raw data for Mg as calculated with PBE in Castep. \label{tab:mg_group_2_pbe_c_raw_dat} } \\ \toprule 
       \multicolumn{1}{l}{$V_{\text{prim}}$ (\AA{}$^3$)} & \multicolumn{1}{l}{CFR fraction} \\ \hline 
\endfirsthead 
\toprule
       \multicolumn{1}{l}{$V_{\text{prim}}$ (\AA{}$^3$)} & \multicolumn{1}{l}{CFR fraction} \\ \hline 
\multicolumn{1}{c}{$\vdots$}& \multicolumn{1}{c}{$\vdots$}\\ 
\endhead 
\multicolumn{1}{c}{$\vdots$}& \multicolumn{1}{c}{$\vdots$}\\ 
\hline \hline 
\endfoot 
\hline \hline 
\endlastfoot
      4.0000 & 0.0000 \\ 
      4.6305 & 0.0000 \\ 
      5.3240 & 0.0000 \\ 
      6.0835 & 0.0000 \\ 
      6.9120 & 0.0000 \\ 
      7.8125 & 0.0000 \\ 
      8.7880 & 0.0000 \\ 
      9.8415 & 0.0000 \\
\end{longtable} 
 
\clearpage

\begin{longtable}{d{4}d{4}} 
  \caption{ Raw data for Ca as calculated with PBE in Castep. \label{tab:ca_group_2_pbe_c_raw_dat} } \\ \toprule 
       \multicolumn{1}{l}{$V_{\text{prim}}$ (\AA{}$^3$)} & \multicolumn{1}{l}{CFR fraction} \\ \hline 
\endfirsthead 
\toprule
       \multicolumn{1}{l}{$V_{\text{prim}}$ (\AA{}$^3$)} & \multicolumn{1}{l}{CFR fraction} \\ \hline 
\multicolumn{1}{c}{$\vdots$}& \multicolumn{1}{c}{$\vdots$}\\ 
\endhead 
\multicolumn{1}{c}{$\vdots$}& \multicolumn{1}{c}{$\vdots$}\\ 
\hline \hline 
\endfoot 
\hline \hline 
\endlastfoot
      4.0000 & 0.0000 \\ 
      4.6305 & 0.0000 \\ 
      5.3240 & 0.0000 \\ 
      6.0835 & 0.0000 \\ 
      6.9120 & 0.0000 \\ 
      7.8125 & 0.0000 \\ 
      8.7880 & 0.0000 \\ 
      9.8415 & 0.0000 \\
\end{longtable} 
 
\clearpage

\begin{longtable}{d{4}d{4}} 
  \caption{ Raw data for Sr as calculated with PBE in Castep. \label{tab:sr_group_2_pbe_c_raw_dat} } \\ \toprule 
       \multicolumn{1}{l}{$V_{\text{prim}}$ (\AA{}$^3$)} & \multicolumn{1}{l}{CFR fraction} \\ \hline 
\endfirsthead 
\toprule
       \multicolumn{1}{l}{$V_{\text{prim}}$ (\AA{}$^3$)} & \multicolumn{1}{l}{CFR fraction} \\ \hline 
\multicolumn{1}{c}{$\vdots$}& \multicolumn{1}{c}{$\vdots$}\\ 
\endhead 
\multicolumn{1}{c}{$\vdots$}& \multicolumn{1}{c}{$\vdots$}\\ 
\hline \hline 
\endfoot 
\hline \hline 
\endlastfoot
      4.0000 & 0.0000 \\ 
      4.6305 & 0.0000 \\ 
      5.3240 & 0.0000 \\ 
      6.0835 & 0.0000 \\ 
      6.9120 & 0.0000 \\ 
      7.8125 & 0.0000 \\ 
      8.7880 & 0.0000 \\ 
      9.8415 & 0.0000 \\
\end{longtable} 
 
\clearpage

\begin{longtable}{d{4}d{4}} 
  \caption{ Raw data for Ba as calculated with PBE in Castep. \label{tab:ba_group_2_pbe_c_raw_dat} } \\ \toprule 
       \multicolumn{1}{l}{$V_{\text{prim}}$ (\AA{}$^3$)} & \multicolumn{1}{l}{CFR fraction} \\ \hline 
\endfirsthead 
\toprule
       \multicolumn{1}{l}{$V_{\text{prim}}$ (\AA{}$^3$)} & \multicolumn{1}{l}{CFR fraction} \\ \hline 
\multicolumn{1}{c}{$\vdots$}& \multicolumn{1}{c}{$\vdots$}\\ 
\endhead 
\multicolumn{1}{c}{$\vdots$}& \multicolumn{1}{c}{$\vdots$}\\ 
\hline \hline 
\endfoot 
\hline \hline 
\endlastfoot
      4.0000 & 0.0000 \\ 
      4.6305 & 0.0000 \\ 
      5.3240 & 0.0000 \\ 
      6.0835 & 0.0000 \\ 
      6.9120 & 0.0000 \\ 
      7.8125 & 0.0000 \\ 
      8.7880 & 0.0000 \\ 
      9.8415 & 0.0000 \\
\end{longtable} 
 
\clearpage

\begin{longtable}{d{4}d{4}} 
  \caption{ Raw data for Ra as calculated with PBE in Castep. \label{tab:ra_group_2_pbe_c_raw_dat} } \\ \toprule 
       \multicolumn{1}{l}{$V_{\text{prim}}$ (\AA{}$^3$)} & \multicolumn{1}{l}{CFR fraction} \\ \hline 
\endfirsthead 
\toprule
       \multicolumn{1}{l}{$V_{\text{prim}}$ (\AA{}$^3$)} & \multicolumn{1}{l}{CFR fraction} \\ \hline 
\multicolumn{1}{c}{$\vdots$}& \multicolumn{1}{c}{$\vdots$}\\ 
\endhead 
\multicolumn{1}{c}{$\vdots$}& \multicolumn{1}{c}{$\vdots$}\\ 
\hline \hline 
\endfoot 
\hline \hline 
\endlastfoot
      4.0000 & 0.0000 \\ 
      4.6305 & 0.0000 \\ 
      5.3240 & 0.0000 \\ 
      6.0835 & 0.0000 \\ 
      6.9120 & 0.0000 \\ 
      7.8125 & 0.0000 \\ 
      8.7880 & 0.0000 \\ 
      9.8415 & 0.0000 \\
\end{longtable} 
 
\clearpage

\clearpage

\onecolumngrid
\section{Raw data for Group 14 elemental solids}

\onecolumngrid
\begin{longtable}{d{4}d{4}d{4}} 
  \caption{ Raw data for C as calculated with PBE in VASP. \label{tab:c_group_14_pbe_v_raw_dat} } \\ \toprule 
       \multicolumn{1}{l}{$V_{\text{prim}}$ (\AA{}$^3$)} & \multicolumn{1}{l}{CFR fraction} & \multicolumn{1}{l}{Band gap from DOS (eV)} \\ \hline 
\endfirsthead 
\toprule
       \multicolumn{1}{l}{$V_{\text{prim}}$ (\AA{}$^3$)} & \multicolumn{1}{l}{CFR fraction} & \multicolumn{1}{l}{Band gap from DOS (eV)} \\ \hline 
\multicolumn{1}{c}{$\vdots$}& \multicolumn{1}{c}{$\vdots$}& \multicolumn{1}{c}{$\vdots$}\\ 
\endhead 
\multicolumn{1}{c}{$\vdots$}& \multicolumn{1}{c}{$\vdots$}& \multicolumn{1}{c}{$\vdots$}\\ 
\hline \hline 
\endfoot 
\hline \hline 
\endlastfoot
      11.3748 & 0.0000 & 4.1395 \\ 
      13.3957 & 0.0000 & 3.7061 \\ 
      15.6427 & 0.0000 & 3.4592 \\ 
      18.1279 & 0.0000 & 3.0474 \\ 
      20.8634 & 0.0212 & 2.3060 \\ 
      23.8610 & 0.1116 & 0.5534 \\ 
      27.1328 & 0.1794 & 0.0000 \\ 
      30.6909 & 0.2407 & 0.0000 \\ 
      34.5471 & 0.3036 & 0.0000 \\ 
      38.7135 & 0.3615 & 0.0000 \\ 
      43.2022 & 0.4113 & 0.0000 \\ 
      48.0250 & 0.4548 & 0.0000 \\ 
      53.1940 & 0.4946 & 0.0000 \\ 
      58.7213 & 0.5302 & 0.0000 \\ 
      64.6187 & 0.5637 & 0.0000 \\ 
      70.8983 & 0.5943 & 0.0000 \\ 
      77.5722 & 0.6228 & 0.0000 \\ 
      84.6522 & 0.6492 & 0.0000 \\ 
      92.1505 & 0.6747 & 0.0000 \\ 
      100.0789 & 0.6986 & 0.0000 \\ 
      108.4495 & 0.7214 & 0.0000 \\ 
      117.2744 & 0.7436 & 0.0000 \\ 
      126.5654 & 0.7654 & 0.0000 \\ 
      136.3346 & 0.7839 & 0.0000 \\ 
      146.5941 & 0.7981 & 0.0000 \\ 
      157.3557 & 0.8104 & 0.0000 \\ 
      168.6315 & 0.8217 & 0.0000 \\ 
      180.4336 & 0.8321 & 0.0000 \\ 
      192.7738 & 0.8421 & 0.0000 \\ 
      205.6642 & 0.8509 & 0.0000 \\ 
      219.1169 & 0.8593 & 0.0000 \\ 
      233.1437 & 0.8669 & 0.0000 \\ 
      247.7567 & 0.8743 & 0.0000 \\ 
\end{longtable} 
 
\clearpage

\begin{longtable}{d{4}d{4}d{4}} 
  \caption{ Raw data for Si as calculated with PBE in VASP. \label{tab:si_group_14_pbe_v_raw_dat} } \\ \toprule 
       \multicolumn{1}{l}{$V_{\text{prim}}$ (\AA{}$^3$)} & \multicolumn{1}{l}{CFR fraction} & \multicolumn{1}{l}{Band gap from DOS (eV)} \\ \hline 
\endfirsthead 
\toprule
       \multicolumn{1}{l}{$V_{\text{prim}}$ (\AA{}$^3$)} & \multicolumn{1}{l}{CFR fraction} & \multicolumn{1}{l}{Band gap from DOS (eV)} \\ \hline 
\multicolumn{1}{c}{$\vdots$}& \multicolumn{1}{c}{$\vdots$}& \multicolumn{1}{c}{$\vdots$}\\ 
\endhead 
\multicolumn{1}{c}{$\vdots$}& \multicolumn{1}{c}{$\vdots$}& \multicolumn{1}{c}{$\vdots$}\\ 
\hline \hline 
\endfoot 
\hline \hline 
\endlastfoot
      40.9168 & 0.0000 & 0.5507 \\ 
      45.5711 & 0.0000 & 0.7547 \\ 
      48.7857 & 0.0064 & 0.8074 \\ 
      50.5655 & 0.0242 & 0.7107 \\ 
      55.9121 & 0.0849 & 0.0000 \\ 
      61.6230 & 0.1457 & 0.0000 \\ 
      67.7100 & 0.1885 & 0.0000 \\ 
      74.1852 & 0.2351 & 0.0000 \\ 
      81.0607 & 0.2779 & 0.0000 \\ 
      88.3483 & 0.3186 & 0.0000 \\ 
      96.0601 & 0.3554 & 0.0000 \\ 
      104.2082 & 0.3890 & 0.0000 \\ 
      112.8044 & 0.4199 & 0.0000 \\ 
      121.8609 & 0.4493 & 0.0000 \\ 
      131.3895 & 0.4764 & 0.0000 \\ 
      141.4023 & 0.5022 & 0.0000 \\ 
      151.9114 & 0.5264 & 0.0000 \\ 
      162.9286 & 0.5491 & 0.0000 \\ 
      174.4660 & 0.5709 & 0.0000 \\ 
      186.5357 & 0.5916 & 0.0000 \\ 
      199.1495 & 0.6114 & 0.0000 \\ 
      212.3195 & 0.6304 & 0.0000 \\ 
      226.0578 & 0.6488 & 0.0000 \\ 
      240.3762 & 0.6661 & 0.0000 \\ 
      255.2868 & 0.6834 & 0.0000 \\ 
\end{longtable}%
 
 
\clearpage

\clearpage

\onecolumngrid
\section{Raw data for Group 18 elemental solids}

\onecolumngrid
%

\end{document}